\documentclass[hidelinks,11pt]{article}

\newcommand{\de}{\,\text{d}}                               
\newcommand{\e}{\operatorname{e}}                               
                              
\newcommand{\ney}{\boldsymbol{y}}                          
\newcommand{\nex}{\boldsymbol{x}}

\newcommand{\p}{\partial}

\newcommand{\R}{\mathbb{R}}

\usepackage{multirow}
\usepackage{amsmath}
\usepackage{amsfonts}
\usepackage{amsmath}
\usepackage{amssymb}  
\usepackage{graphicx}
\usepackage{caption}
\usepackage{subcaption}
\usepackage{hyperref}

\title{\bf Windowed Green Function method\\ for layered-media
  scattering }

\author
       {Oscar P. Bruno$^1\footnote{Corresponding author: \texttt{obruno@caltech.edu}.}$, Mark~Lyon$^2$, Carlos P\'erez-Arancibia$^{1}$ and
       Catalin~Turc$^3$\\ \\
       \small{$^1$Computing \& Mathematical Sciences, California Institute of Technology}\smallskip\\
       \small{$^2$Department of Mathematics and Statistics, University of New Hampshire}\smallskip\\
      \small{ $^3$Department of Mathematics, New Jersey Institute of Technology }}
       
\date{}
\date{\today}

\begin{document}
\maketitle


\begin{abstract}
  This paper introduces a new Windowed Green Function (WGF) method for
  the numerical integral-equation solution of problems of
  electromagnetic scattering by obstacles in presence of dielectric or
  conducting half-planes. The WGF method, which is based on use of
  smooth windowing functions and integral kernels that can be
  expressed directly in terms of the free-space Green function, does
  not require evaluation of expensive Sommerfeld integrals. The
  proposed approach is fast, accurate, flexible and easy to
  implement. In particular, straightforward modifications of existing
  (accelerated or unaccelerated) solvers suffice to incorporate the
  WGF capability. The mathematical basis of the method is simple: the
  method relies on a certain integral equation posed on the union of
  the boundary of the obstacle and a small flat section of the
  interface between the penetrable media. Numerical experiments
  demonstrate that both the near- and far-field errors resulting from
  the proposed approach decrease faster than any negative power of the
  window size. In the examples considered in this paper the proposed
  method is up to thousands of times faster, for a given accuracy,
  than a corresponding method based on the layer-Green-function.
\end{abstract}

\section{\label{sec:introduction}Introduction}
The solution of problems of scattering by obstacles or defects in
presence of planar layered dielectric or conducting media has
typically required use of Sommerfeld integrals and associated layer
Green functions---which automatically enforce the relevant
transmission conditions on the unbounded flat surfaces and thus reduce
the scattering problems to integral equations on the obstacles and/or
defects. As is well known, however, the numerical evaluation of layer
Green functions and their derivatives, which amounts to computation of
certain challenging Fourier
integrals~\cite{Chew1995waves,Sommerfeld1909}, are extremely expensive
and give rise to a significant bottleneck in layer-media simulations
(see e.g.~\cite{Cai:2002vt} for details). This paper presents a novel
integral-equation approach for problems involving layered media. The
new approach, which is based on use of certain ``windowing'' functions
and considerations associated with the method of stationary phase,
\emph{does not require use of expensive Sommerfeld integrals}.
Numerical experiments demonstrate that both the near- and far-field
errors resulting from the proposed approach decrease faster than any
negative power of the window size.

A variety of methods have been provided for the solution of problems
of scattering by obstacles in presence of layered media.  Amongst the
most effective such approaches we mention 1)~Methods which evaluate
Sommerfeld integrals on the basis of path-integration in the complex
plane~\cite{Paulus:2000vr,Cai:2000bl,Cai:2002vt,PerezArancibia:2014fg}
(such approaches require numerical evaluation of integrals of
functions that oscillate, grow exponentially in a bounded section of
the integration path and, depending on the relative position of the
source and observation points to the interface between the two media,
may decay slowly at infinity); 2)~The complex images method reviewed
in~\cite{Aksun:2009fn} (a discussion indicating certain instabilities
and inefficiencies in this method is presented
in~\cite[section~5.5]{Cai:2000bl}); and 3)~The steepest descent
method~\cite{Cui:1998fw,Cui:1999tb} which, provided the steepest
descent path is known, reduces the Sommerfeld integral to an integral
of an exponentially decaying function (unfortunately, however, the
determination of steepest descent paths for each observation point can
be challenging and expensive). As is well known, in any case, all of
these methods entail significant computational
costs~\cite{Cai:2002vt}.

The approach proposed in this paper bears similarities with certain
``finite-section'' methods in the field of rough-surface
scattering. These methods utilize approximations based on truncated
portions of a given unbounded rough
surface~\cite{Meier:2001kj,Zhao:2005ue,Saillard:2011jj} and, in some
cases, they incorporate a
``taper''~\cite{Zhao:2005ue,Spiga:2008co,Miret:2014jg} to eliminate
artificial reflections from the edges of the finite sections. In fact
the smooth taper function utilized in~\cite{Miret:2014jg} (Figure~2 in
that reference) resembles the smooth windowing function we use
(Figure~\ref{fig:window_function} below and
reference~\cite{Bruno:2014cf}). But as indicated in comments provided
in section~\ref{sec:win_alg} below in regards to certain slow-rise
windowing functions, essential differences exist between the
finite-section approaches and the methods proposed in this paper. In
particular, with exception of the slow-rise windowing function
method~\cite{Bruno:2014cf,MonroJr:2008te}, none of the previous
tapered rough surface algorithms has demonstrated high-order
convergence as the width of the finite sections tend to infinity.

In section~\ref{sec:num_exp} the proposed WGF method is compared
against the high-order integral equation method recently introduced
in~\cite{PerezArancibia:2014fg}, which is based on the accurate and
efficient evaluation of the Sommerfeld integrals. In the examples
considered in that section the proposed method is up to thousands of
times faster, for a given accuracy, than a corresponding method based
on the layer-Green-function. A much larger improvement in the
computational cost is expected for problems of electromagnetic
scattering by defects and obstacles in multi-layer structures in two-
and three-dimensional spaces, which will be addressed in future
contributions.

The proposed methodology is presented in sections~\ref{sec:win_alg}
and~\ref{field_eval}.  A variety of numerical results presented in
sections~\ref{sec:win_alg} and~\ref{sec:num_exp} demonstrate the
accuracy and speed of the proposed approach.

\section{Windowed Green Function Method}\label{sec:win_alg}

\begin{figure}[ht!]
  \centering
    \includegraphics[scale=0.9]{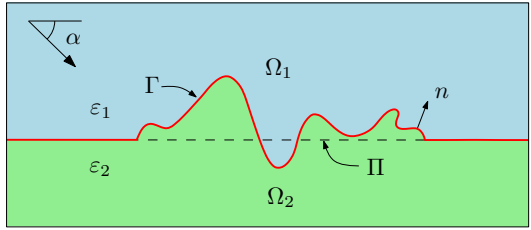}
    \caption{Description of the problem under consideration:
      scattering by a defect in a dielectric or conducting
      plane. $\Gamma$ denotes the interface between the two media
      while $\Pi$ denotes the interface between the upper- and
      lower-half planes.}
  \label{fig:geometry}
\end{figure}

We consider two-dimensional TE and TM polarized dielectric
transmission problems. As is well known, the $z$ components $u=E_z$
and $u=H_z$ of the total electric and magnetic fields satisfy the
Helmholtz equation $\Delta u+k_j^2u=0$ in $\Omega_j$, $j=1,2$ (see
Figure~\ref{fig:geometry}), where, letting $\omega>0$, $\varepsilon_j
>0$, $\mu_0 >0$, and $\sigma_j \geq 0$ denote the angular frequency,
the electric permittivity, the magnetic permeability of vacuum, and
the electrical conductivity, the wavenumber $k_j$ is defined by
$k_j^2=\omega^2(\varepsilon_j+i\sigma_j/\omega)\mu_0$, $j=1,2$.  In
either case the total field is given by
\begin{equation}\label{eq:rep_0}
u = \left\{\begin{array}{ccc}
u_1+u^{\mathrm{inc}}&\mbox{in}& \Omega_1,\\
u_2&\mbox{in}& \Omega_2,
\end{array}\right.
\end{equation}
where denoting by $\alpha\in (-\pi,0)$ the incidence angle measured
from the horizontal (see Figure~\ref{fig:geometry}),
$u^{\mathrm{inc}}(\nex)= \e^{ik_1(x_1\cos\alpha +x_2\sin\alpha)}$,
$u_1$ and $u_2$ denote the incident plane-wave and the reflected and
transmitted waves, respectively. As is known (see
e.g.~\cite{DeSanto:1997es}), the scattered and transmitted fields
$u_1$ and $u_2$ admit the representations
\begin{subequations}
\begin{eqnarray}
u_1 &=& \mathcal D_1\left[\varphi\right]-\mathcal S_1\left[\psi\right]\quad\mbox{ in }\quad\Omega_1,\label{eq:rep_1}\\
u_2 &=& -\mathcal D_2\left[\varphi\right]+\mathcal S_2\left[\psi\right]\quad\mbox{ in }\quad\Omega_2,\label{eq:rep_i}
\label{eq:rep_2}
\end{eqnarray}
\label{eq:representation}\end{subequations}
in terms of the total field $\varphi=u|_{\Gamma}$ and its normal
derivative $\psi=\frac{\p u}{\p n}$ on $\Gamma$, where letting
$G_j(\nex,\ney) = iH_0^{(1)}(k_j|\nex-\ney|)/4$, $j=1,2$ denote the
free-space Green function for the Helmholtz equation with wavenumber
$k_j$, the single- and double-layer potentials in
equation~\eqref{eq:representation} are defined by
\begin{equation}\begin{split}
  \mathcal S_j[\eta](\nex) &= \int_{\Gamma}G_j(\nex,\ney)\eta(\ney)\de s_{\ney},\quad\mbox{and}\\
  \mathcal D_j[\eta](\nex) &= \int_{\Gamma}\frac{\p G_j}{\p n_{\ney}}(\nex,\ney)\eta(\ney)\de  s_{\ney},
  \end{split}\label{eq:pot_gamma}
\end{equation}
respectively.  By evaluating the fields~\eqref{eq:representation} and
their normal derivatives on $\Gamma$ and using the transmission
conditions
$$ u_2-u_1=u^{\mathrm{inc}},\quad \nu\frac{\p u_2}{\p n}-\frac{\p u_1}{\p n}=\frac{\p u^{\mathrm{inc}}}{\p n}\mbox{ on }\Gamma,
$$
(with $\nu = 1$ and $\nu=\varepsilon_1/\varepsilon_2$ in TE- and
TM-polarizations respectively) we obtain the second-kind system of
integral equations~\cite{Kittappa:1975vr}
\begin{equation}\label{eq:int_eq_full}
E\phi + T\phi =\phi^{\mathrm{inc}} \quad \mbox{on}\quad \Gamma
\end{equation}
for the surface currents $\phi$, where $$ E = \left[\begin{array}{cc}
    1& 0\\
    0& \frac{1+\nu}{2}
	\end{array}\right],\quad \phi = \left[\begin{array}{cc}
u|_\Gamma\\
\frac{\p u}{\p n}|_{\Gamma}
	\end{array}\right],\quad \phi^{\mathrm{inc}} = \left[\begin{array}{cc}
u^{\mathrm{inc}}|_\Gamma\\
\frac{\p u^{\mathrm{inc}}}{\p n}|_{\Gamma}
	\end{array}\right],
$$ and where
\begin{equation}
T = \left[\begin{array}{cc}
	D_2-D_1& -\nu S_2+ S_1\\
	N_2-N_1& -\nu K_2+ K_1\\
	\end{array}\right]\label{eq:transmission_operator}
\end{equation}
is defined in terms of the boundary integral operators defined by the
expressions $S_j[\eta](\nex)$ and $D_j[\eta](\nex)$ as well as
\begin{equation*}
  N_j[\eta](\nex) = \frac{\p \mathcal D_j\eta}{\p n}(\nex)\quad\mbox{and}\quad
  K_j[\eta](\nex) = \displaystyle\int_{\Gamma}\frac{\p G_j}{\p n_{\nex}}(\nex,\ney)\eta(\ney)\de s_{\ney}
\end{equation*}
for $\nex\in \Gamma$ and for $j=1$, 2.

\begin{figure}[ht!]
  \centering
    \includegraphics[scale=0.9]{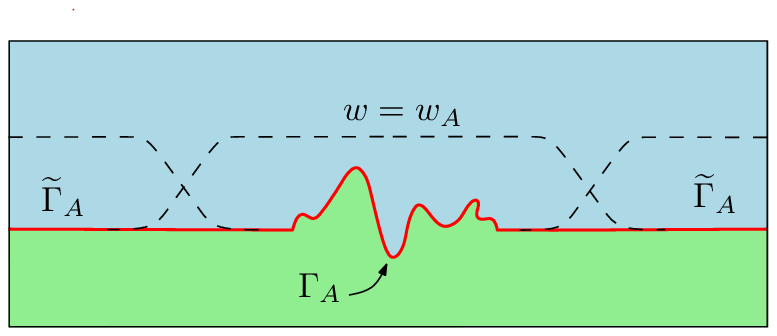}
    \caption{Window function $w=w_A$ and the windowed sections
      $\Gamma_A$ and $\widetilde\Gamma_A$ of the unbounded curve~$\Gamma$.}
  \label{fig:window_function}
\end{figure}

Instead of solving the problem on the entire infinite plane a locally
windowed problem could be used in an attempt to obtain local currents
over relevant portions of the geometry. To pursue this idea we
introduce a smooth windowing function $w_A$ (which is depicted
in~Figure~\ref{fig:window_function}) which is non-zero in an interval
of length $2A$, and which has a slow rise: $w_A(x_1) = f(x_1/A)$ for
some fixed window function $f$.  (Note that, with such a definition,
$w_A$ rises from zero to one in a region of length proportional to
$A$; see~\cite{Bruno:2014cf,MonroJr:2008te}. As demonstrated in those
references, the slow rise of the window function is essential to
ensure fast convergence of the approximation.)  For notational
simplicity, the subindex $A$ will be dropped in what follows, and we
will thus write $w(x_1)$ instead of $w_A(x_1)$. The parts of the
boundary $\Gamma$ where $w(x_1)\neq 0$ and $\widetilde
w(x_1)=1-w(x_1)\neq 0$, further, will be denoted by $\Gamma_A$ and
$\widetilde\Gamma_A$, respectively. The width~$2A>0$ of the support of
the window function~$w$ is selected in such a way that $\widetilde
w(x_1)$ vanishes on any corrugations that exist on the surface
$\Gamma$, as well as on any additional obstacles that may exist above
and/or below $\Gamma$. (For notational simplicity our derivations are
presented for cases for which the corrugations on the surface $\Gamma$
are the only departures from planarity, but, as demonstrated by
Figure~\ref{fig:CKnear}, our algorithms are also applicable in cases
in which additional scatterers exist.)

Utilizing the windowing function $w$ and letting $W=w\cdot I$, where
$I$ is the $2\times 2$ identity matrix, we consider the preliminary
approximate equation
\begin{equation}\label{eq:angledep}
  E\phi^{
    \star} +  T W \phi^{
    \star}  = \phi^{\mathrm{inc}}\quad \mbox{on}\quad \Gamma_A
\end{equation}
(where the new unknown $\phi^{ \star}$ is defined on $\Gamma_A$), and,
in order to assess the errors inherent in this approximation, the form
\begin{equation}\label{eq:exact}
E \phi+  TW\phi  = \phi^{\mathrm{inc}} -  T(I-W) \phi \quad\mbox{on}\quad \Gamma_A
\end{equation}
of the exact equation~\eqref{eq:int_eq_full}.  Using
integration-by-parts and employing the method of stationary-phase, it
follows~\cite{Bruno-PA} that the term $T(I-W)\phi$ is
super-algebraically small (i.e., smaller than $C_p(kA)^{-p}$ for any
positive integer $p$ as $kA\to\infty$, where $C_p$ is a $p$-dependent
constant) in the region $\{w=1\}$, and, thus, as shown
in~\cite{Bruno-PA}, that the solution~$\phi^{ \star}$
of~\eqref{eq:angledep} is a highly accurate approximation of $\phi$
throughout the center region $\{w=1\}$ of the surface $\Gamma_A$
provided $A$ is large enough. However, it is easy to see that, to
correctly take into account fields reflected from the planar portions
of the surface, the needed window sizes may be very large---especially
so for incidence angles approaching grazing.  

To demonstrate this fact we use equation~\eqref{eq:angledep} to
approximate the solution of the TE problem of scattering of a
plane-wave by a semi-circular bump of radius $a=1$ placed directly on
top of a planar dielectric surface. The problem was discretized using
a graded mesh over the surface of the bump and on the windowed portion
of the planar interface, on the basis of a direct generalization of
the Nystr\"om method presented in~\cite{Kress:1990vm} with $p=3$.  For
this example the wavenumbers $k_1$ and $k_2$ in the regions above and
below the plane were set to $4\pi$ and $8\pi$, respectively, and
approximately 20 points per unit length of the surface of the bump and
the surrounding were used.

\begin{figure}[ht!]
  \centering
 \includegraphics[scale=0.35]{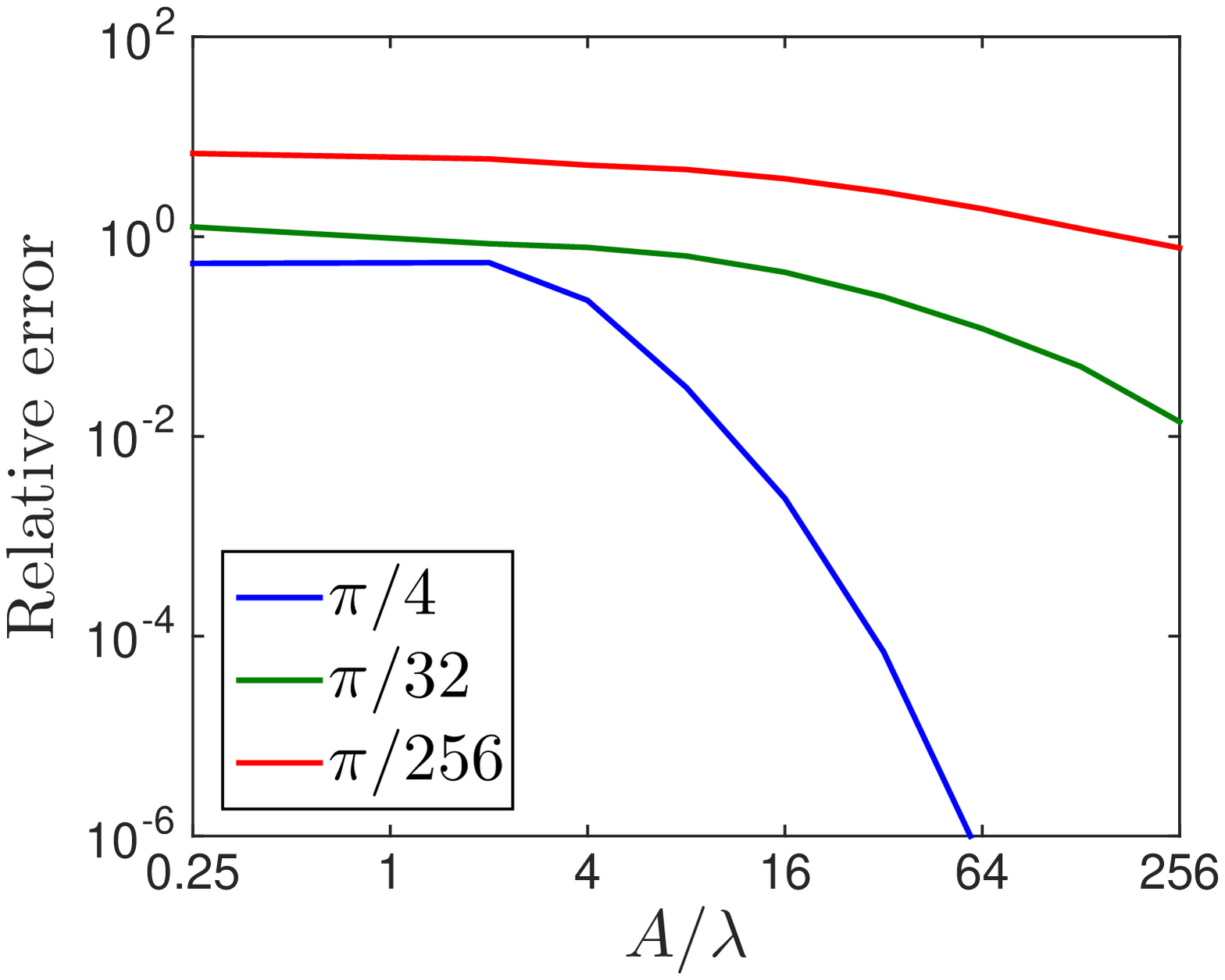}\hspace{0.5cm}
 \includegraphics[scale=0.35]{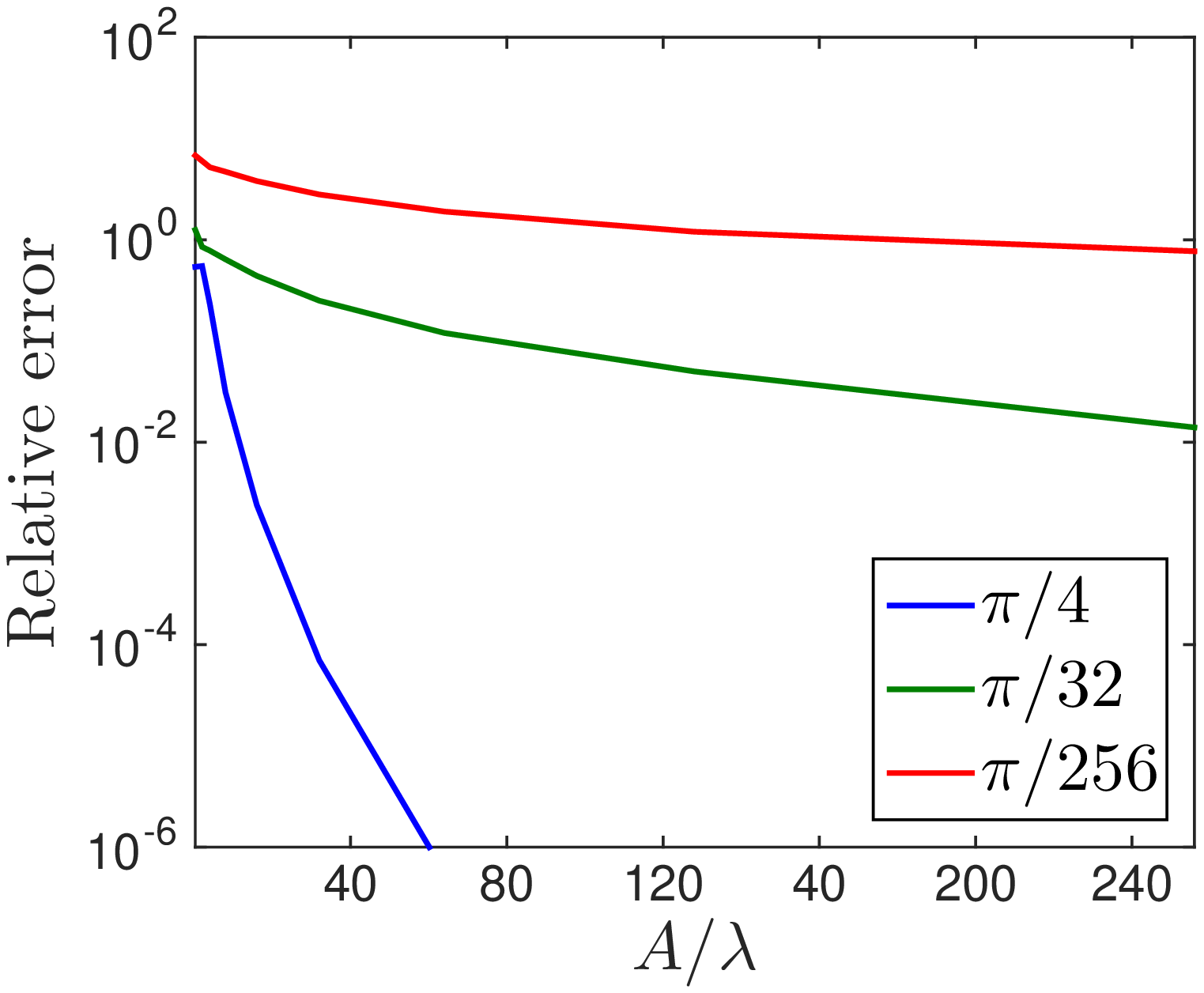}
 \caption{Errors in the integral densities resulting from numerical
   solution of~\eqref{eq:angledep} by means of a naive
   implementation of the WGF method for a semi-circular bump-shaped
   defect, for various window sizes and angles of incidence. Left:
   log-log scale. Right: semi-log scale. Clearly, the window size
   required by the naive method to produce a given accuracy increases
   dramatically as the angle of incidence approaches grazing.}
  \label{fig:naive_WGFM}
\end{figure}

As shown in Figure~\ref{fig:naive_WGFM}, the naive windowing approach
embodied in~\eqref{eq:angledep} requires large regions of the planar
interface to be discretized as the incidence angle decreases. For
accurate calculations at even moderate angles, a large number of
wavelengths must be present in the window region, well beyond the
extent of the non-planar local geometry.

\begin{figure}[ht!]  
  \centering
    \includegraphics[scale=0.9]{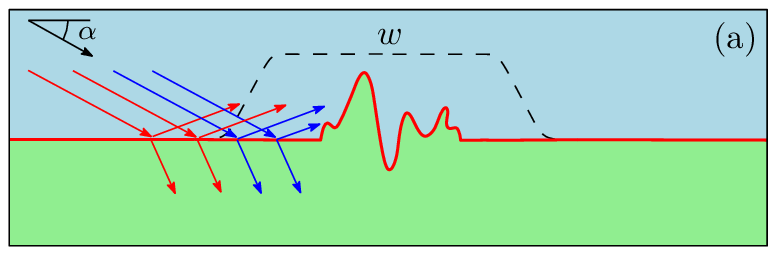}\hspace{0.5cm}
        \includegraphics[scale=0.9]{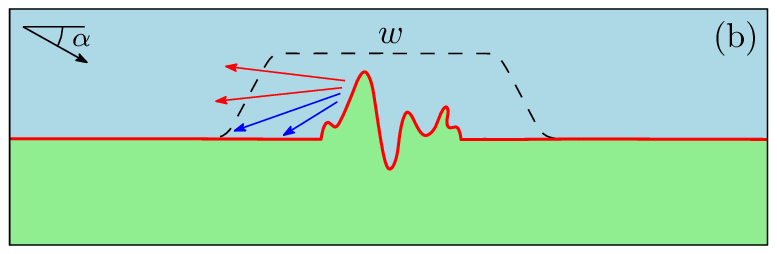}
        \caption{Physical elements underlying the WGF method.}
        \label{fig:physical_interpretation}
\end{figure}

In order to provide an insight into the source of the errors displayed
in Figure~\ref{fig:naive_WGFM} we present
Figure~\ref{fig:physical_interpretation}. Figure~\ref{fig:physical_interpretation}(a)
presents rays incident on the left planar region as well as their
reflection and transmission. Clearly, in view of the incidence angle
considered these reflected fields subsequently illuminate the
defect. The blue rays, for example, represent the reflections that are
correctly taken into account in the solution of
equation~\eqref{eq:angledep} (since they lie within the windowed region),
while the red arrows represent reflections that are
neglected. Figure~\ref{fig:physical_interpretation}(b), on the other
hand, represents reflections by the defect. The color-code in the left
figure carries over to the right figure: the blue (resp. red) rays in
Figure~\ref{fig:physical_interpretation}(b) represent the fields
scattered by the defect which arise from the blue (resp. red) arrows
in Figure~\ref{fig:physical_interpretation}(a). We remark that the
scattering of the field represented by the red arrows is not taken
into account by~\eqref{eq:angledep}, which gives rise to the errors
observed in Figure~\ref{fig:naive_WGFM}. We also note that the
relatively fast convergence demonstrated by the blue curves in
Figure~\ref{fig:naive_WGFM} is explained by the fact that for near
normal incidence ($\alpha\approx-\pi/2$) there is not much ``red
field" interacting with the defect. In contrast, for incidence near
grazing ($\alpha\approx 0$), ``red fields" from regions far away from
the windowed area do interact with the defect. This explains the poor
convergence properties demonstrated by the green and red curves in
Figure~\ref{fig:naive_WGFM}: the fields neglected in the naive
approach give rise to important contributions as $\alpha$ decreases.

To address this difficulty we consider again the exact integral
equation~\eqref{eq:exact} and we substitute the unknown density~$\phi$
on the right-hand side of this equation by the corresponding (known)
density~$\phi^f$ associated with the problems of scattering and
transmission of a plane-wave by a perfectly flat infinite plane. Since
a superalgebraically small portion of the field reflected by the
windowed region reflects back into the windowed region upon reflection
from the plane outside the windowed region, we conclude that the error
arising from the substitution of $\phi$ by $\phi^f$ results in
superalgebraically small errors in equation~\eqref{eq:exact}
throughout the region $\{w=1\}$. We thus obtain the approximate
equation
\begin{equation}
E \phi^w+  T W\phi^w =  \phi^{\mathrm{inc}} -  T (I-W) \phi^f\quad\mbox{on}\quad \Gamma_A,\label{eq:IE_1st_vesion}
\end{equation}
whose solution $\phi^w$ is a superalgebraically close approximation of
the exact solution $\phi$ throughout the region $\{w=1\}$. In order to
evaluate the term $T(I-W) \phi^f$ we note that since $(I-W)\phi^f$ is
zero everywhere $\Gamma_A$ deviates from the planar
boundary~$\Pi=\{(x_1,x_2)\in\R^2:x_2=0\}$ (depicted in
Figure~\ref{fig:geometry}), we have
  $$T (I-W) \phi^f = T_{\Pi}(I-W)\phi^f,$$ 
  where letting the layer potentials~$\mathcal S^\Pi_j$ and $\mathcal D_j^\Pi$ be given by
\begin{equation}\begin{split}
  \mathcal S^\Pi_j[\eta](\nex) &= \int_{\Pi}G_j(\nex,\ney)\eta(\ney)\de s_{\ney},\quad\mbox{and}\\
  \mathcal D^\Pi_j[\eta](\nex) &= \int_{\Pi}\frac{\p G_j}{\p n_{\ney}}(\nex,\ney)\eta(\ney)\de  s_{\ney},
  \end{split}\label{eq:pot_plane}
\end{equation}
the operator~$T_\Pi$ is defined as 
\begin{equation*}
T_\Pi = \left[\begin{array}{cc}
	D^\Pi_2-D^\Pi_1& -\nu S^\Pi_2+ S^\Pi_1\\
	N^\Pi_2-N^\Pi_1& -\nu K^\Pi_2+ K^\Pi_1\\
	\end{array}\right]\label{eq:transmission_operator_Pi}
\end{equation*}
in terms of the boundary integral operators defined by the expressions
$S^\Pi_j[\eta](\nex)$ and $D^\Pi_j[\eta](\nex)$ as well as 
\begin{equation*}
 N^\Pi_j[\eta](\nex) = \frac{\p \mathcal D^\Pi_j\eta}{\p n}(\nex)\big|_{\Gamma}\quad\mbox{and}\quad
K^\Pi_j[\eta](\nex) = \displaystyle\int_{\Pi}\frac{\p G_j}{\p n_{\nex}}(\nex,\ney)\eta(\ney)\de s_{\ney}
\end{equation*}
for $\nex\in\Gamma$ and for $j=1$, 2.  Thus
equation~\eqref{eq:IE_1st_vesion} becomes
\begin{equation}
E \phi^w+  TW \phi^w = \phi^{\mathrm{inc}} -  T_\Pi \phi^f + T_\Pi W \phi^f\quad\mbox{on}\quad \Gamma_A.\label{eq:IE_2nd_version}
\end{equation}
Clearly the expression $T_{\Pi} W \phi^f$ can be evaluated by means of
integration on the bounded region
$\Pi\cap\{(x_1,x_2)\in\R^2:w(x_1)\neq 0\}$, and the expression
$T_{\Pi} \phi^f$ can be computed in closed form:
\begin{equation}
T_{\Pi} \phi^f = \left\{\begin{array}{ccl}
\displaystyle\left[u^{\mathrm{inc}}-u^{f},\frac{\p(u^{\mathrm{inc}}- u^{f})}{\p n}\right]^T&\mbox{on}&\Gamma\setminus \Pi,\\
\displaystyle\left[u^{\mathrm{inc}}-u^{f},\frac{\p(u^{\mathrm{inc}}-(1+\nu)u^{f}/2 )}{\p n}\right]^T&\mbox{on}&\Gamma\cap\Pi,
\end{array}\right.\label{eq:right_hand_side}
\end{equation}
where~$u^f$ is the total field resulting from the solution of the
problem of scattering by the flat dielectric plane with
boundary~$\Pi$~\cite[Chapter 2]{Chew1995waves}.

\begin{figure}[ht!]
  \centering
 \includegraphics[scale=0.35]{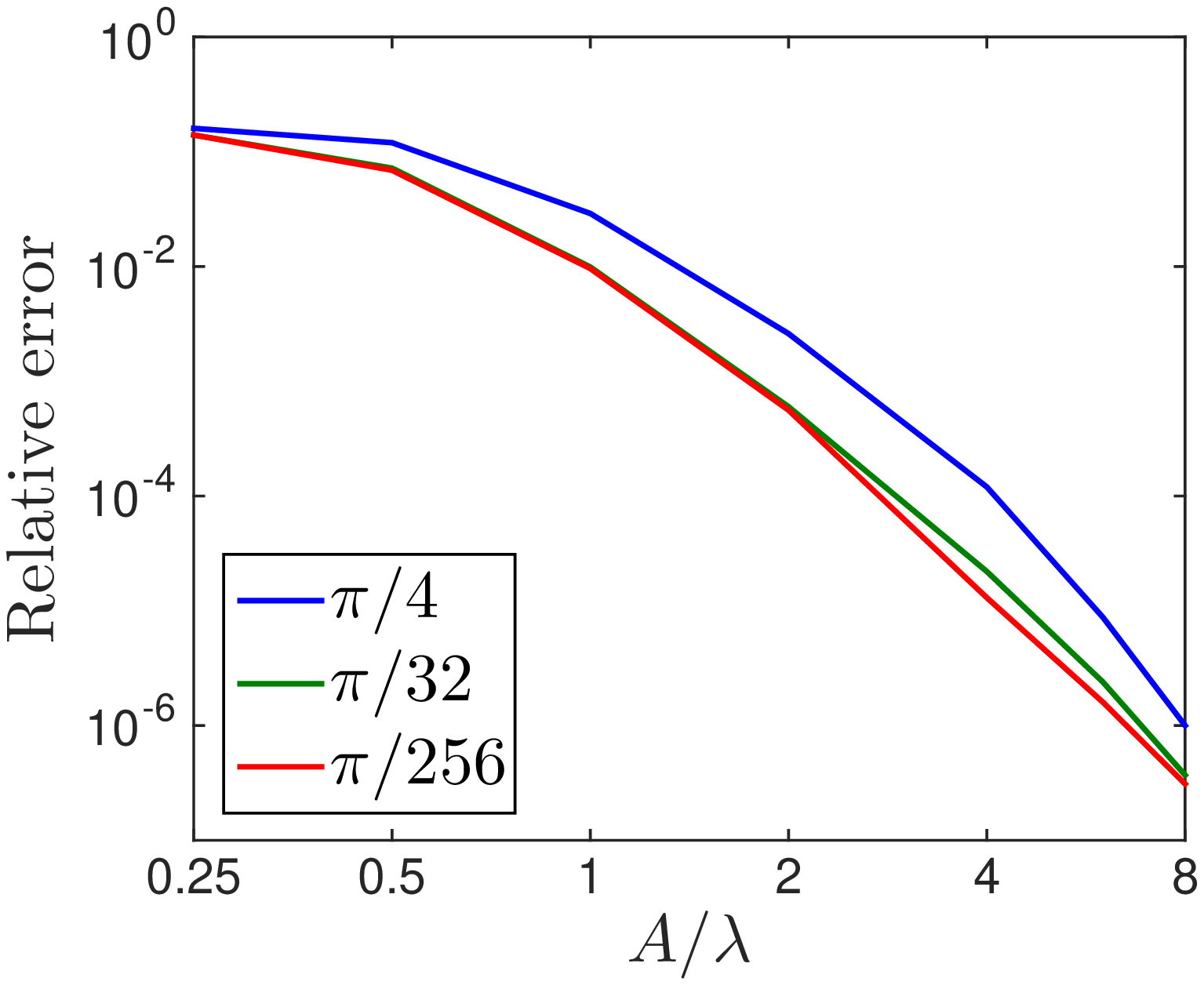}\hspace{0.5cm}
 \includegraphics[scale=0.35]{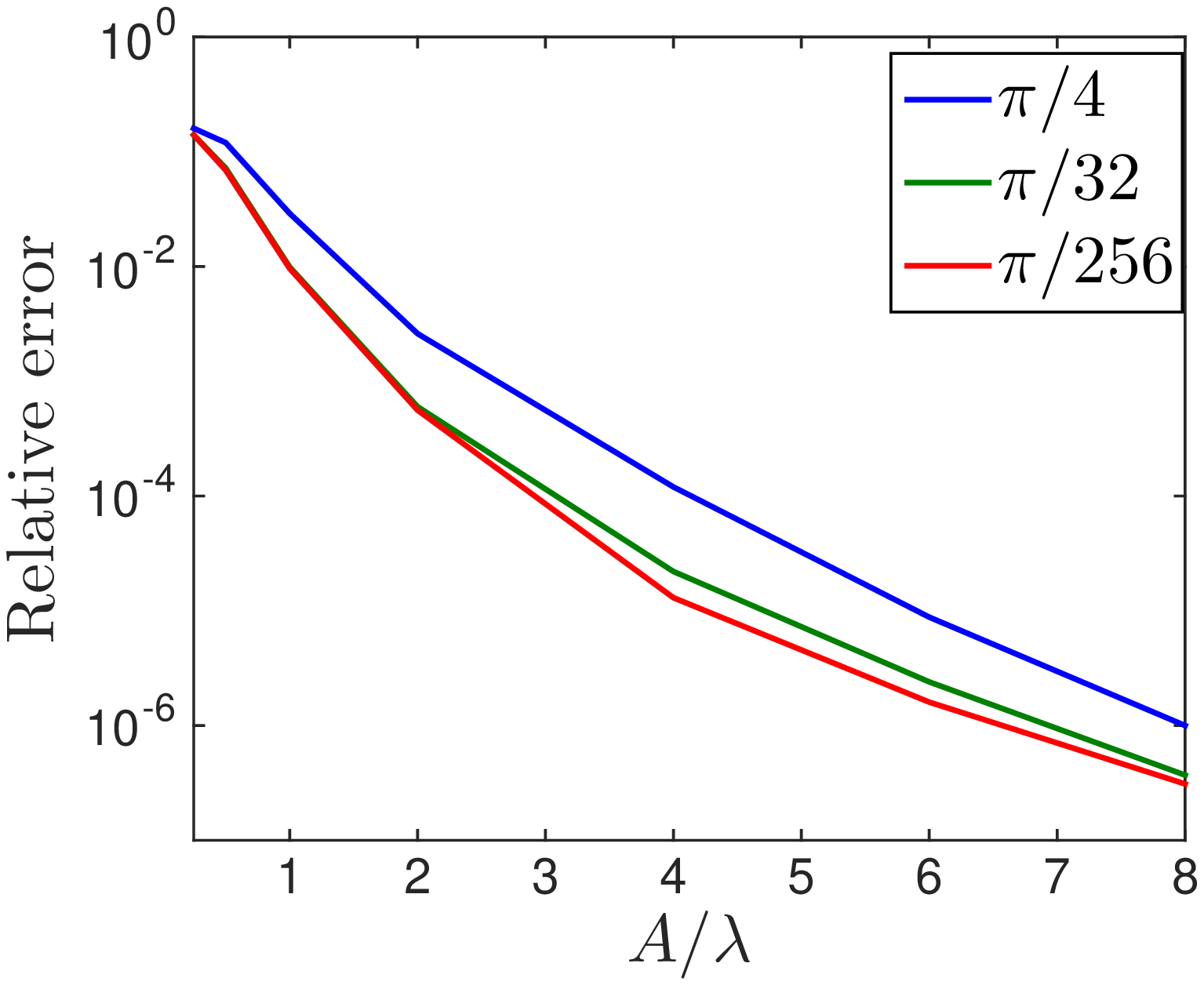}
 \caption{Errors in the integral densities $\phi^w$ on the surface of
   the defect resulting from numerical solution
   of~\eqref{eq:IE_2nd_version}, for a semi-circular bump-shaped
   defect, and for various window sizes and angles of
   incidence---including extremely shallow incidences. Left: log-log
   scale. Right: semi-log scale.  Clearly, this version of the WGF
   method computes integral densities with super-algebraically high
   accuracy uniformly for all angles of incidence
   (cf. Figure~\ref{fig:naive_WGFM}).}
  \label{fig:window_method_only}
\end{figure}

From the discussion above we see that, on the set $\{w=1\}$, the
(superalgebraically high) accuracy of the solution $\phi^w$
of~\eqref{eq:IE_2nd_version} does not deteriorate as the incidence
angle $\alpha$ tends to zero.  As shown in section~\ref{field_eval}
below, further, the solution $\phi^w$ can be used to produce the total
field~$u$ everywhere in space as well as the associated far field
pattern. To conclude this section, in
Figure~\ref{fig:window_method_only} we demonstrate the fast and
angle-independent convergence of $\phi^w$ to $\phi$: clearly the value
of $A$ required to obtain an accurate approximation of the exact
solution has been reduced substantially and the errors are uniformly
small as the incidence angle decreases to zero.

\section{Field evaluation\label{field_eval}}

An analysis similar to the one presented in section~\ref{sec:win_alg}
for the density $\phi^w = [\varphi^w,\psi^w]^T$ shows that
substitution of $\phi=[\varphi,\psi]^T$ by
$[w\varphi^w+(1-w)\varphi^{f},w\psi^w+(1-w)\psi^{f}]^T$ in
\eqref{eq:representation} produces the fields $u_1$ and $u_2$ with
superalgebraically high accuracy in a neighborhood of the region
$\{w=1\}$ in $\R^2$, and, in particular, on a closed disc $D$ such as
the one depicted in Figure~\ref{fig:far_field_depic}.  After some
manipulations similar to those presented in the derivation of
\eqref{eq:right_hand_side} above, the resulting formula can be
re-expressed into a formula for the total field in terms of surface
potentials defined on both $\Gamma$ and $\Pi$, namely
\begin{subequations}
 \begin{equation}
 \begin{split}
u(\nex)=&\  \mathcal D_1\left[w\varphi^w\right](\nex)-\mathcal S_1\left[w\psi^w\right](\nex) -\mathcal D^\Pi_{1}\left[w\varphi^f\right](\nex)+\mathcal S^\Pi_{1}\left[w\psi^f\right](\nex) \\
& +\left\{\begin{array}{ll}
u^f(\nex),&\nex\in\{x_2\geq 0\},\medskip\\
0,&\nex\in\{x_2<0\}
\end{array}\right.\end{split}
\end{equation}
for $\nex\in\Omega_1$, and
 \begin{equation}\begin{split}
u(\nex) =&\ -\mathcal D_2\left[w\varphi^w\right](\nex)+\mathcal S_2\left[\nu w\psi^w\right](\nex) +\mathcal D^\Pi_{2}\left[w\varphi^f\right](\nex)-\mathcal S^\Pi_{2}\left[\nu w\psi^f\right](\nex)\\  &+\left\{\begin{array}{ll}
0,&\nex\in\{x_2\geq 0\},\medskip\\
u^f(\nex),&\nex\in\{x_2<0\} \end{array}\right.\end{split}\end{equation}\label{eq:evaluation_formulae}\end{subequations}
for~$\nex\in\Omega_2$. 

\begin{figure}[h!]
    \centering
    \begin{subfigure}[b]{0.3\textwidth}
        \centering
        \includegraphics[scale=0.45]{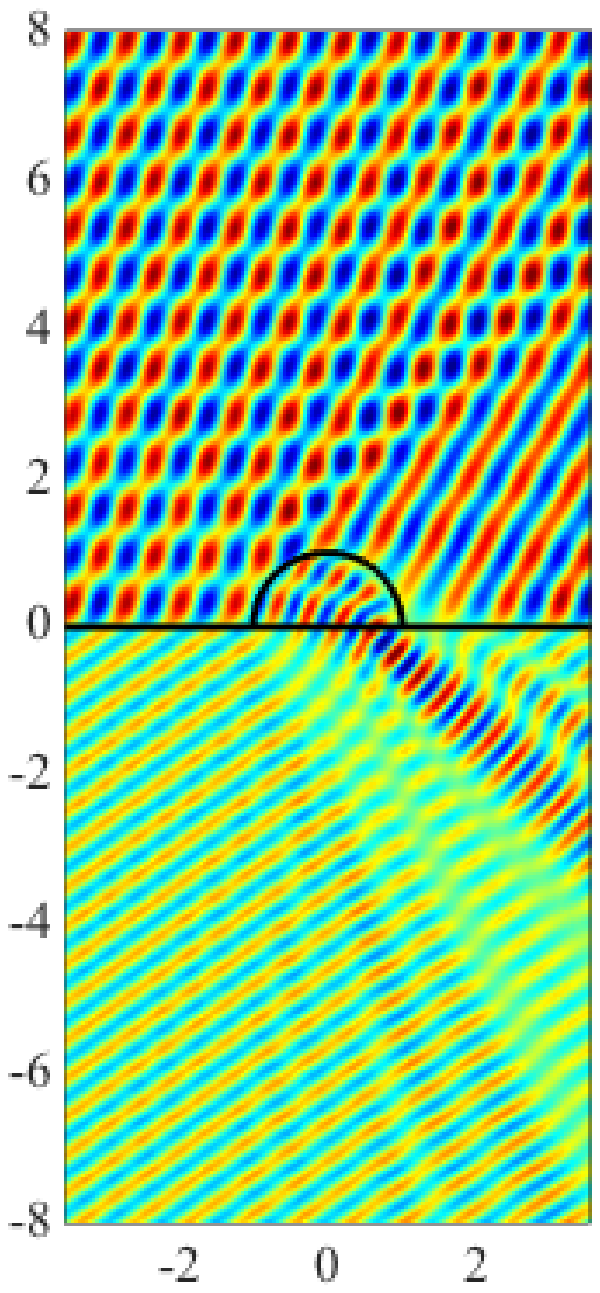}
        \caption{WGF method.}
        \label{fig:WGFM_pi_2}
    \end{subfigure}
    \hfill
    \begin{subfigure}[b]{0.3\textwidth}
        \centering
        \includegraphics[scale=0.45]{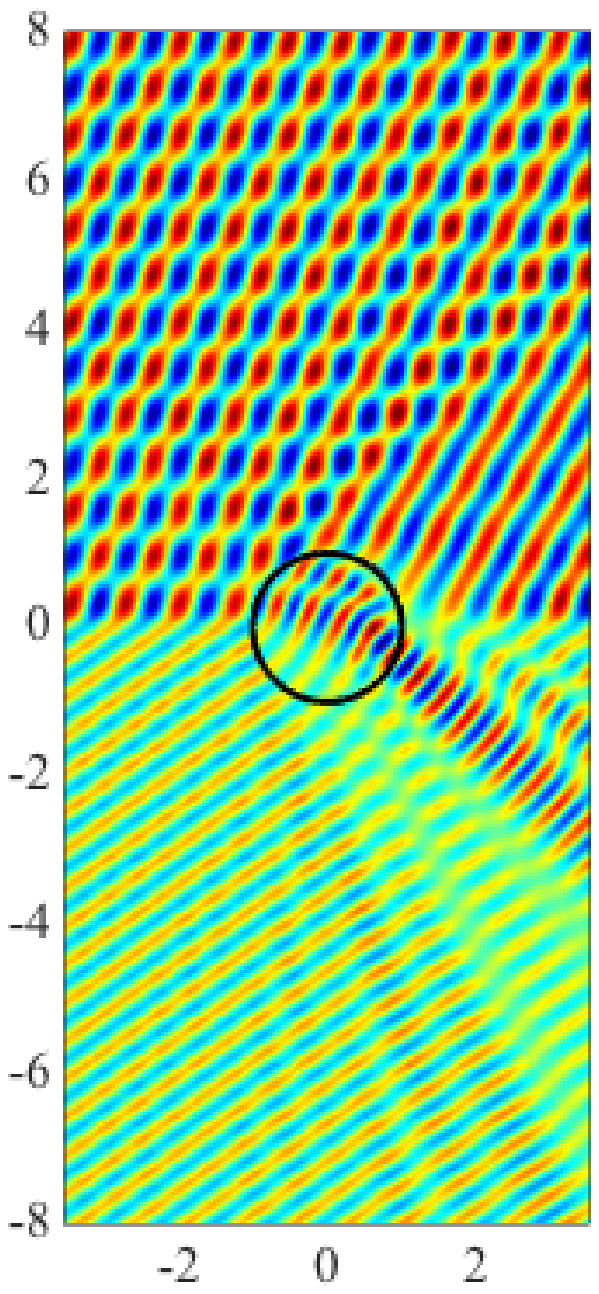}
        \caption{LGF method.}
        \label{fig:LGFM_pi_2}
    \end{subfigure}
    \hfill
    \begin{subfigure}[b]{0.3\textwidth}
        \centering
        \includegraphics[scale=0.45]{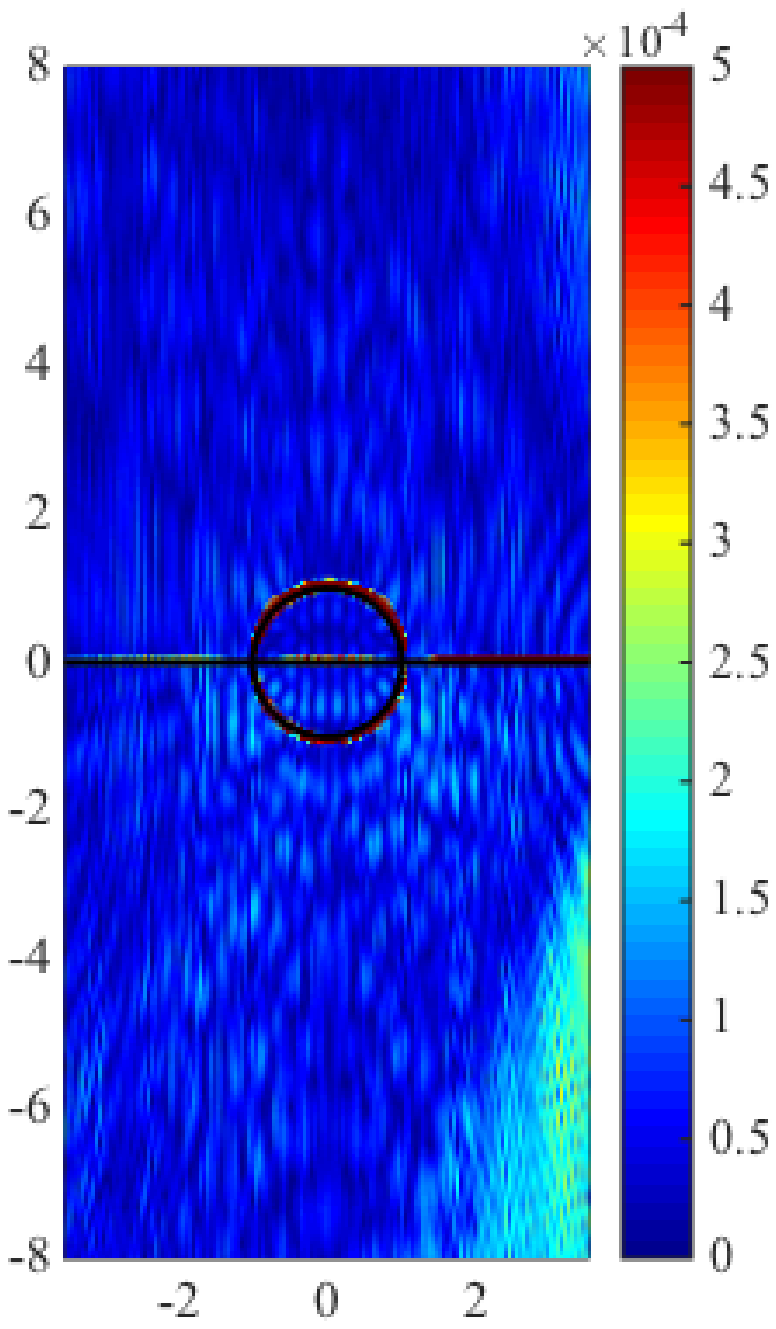}
        \caption{Difference.}
        \label{fig:pi_2_error}
    \end{subfigure}\\
     \begin{subfigure}[b]{0.3\textwidth}
        \centering
        \includegraphics[scale=0.45]{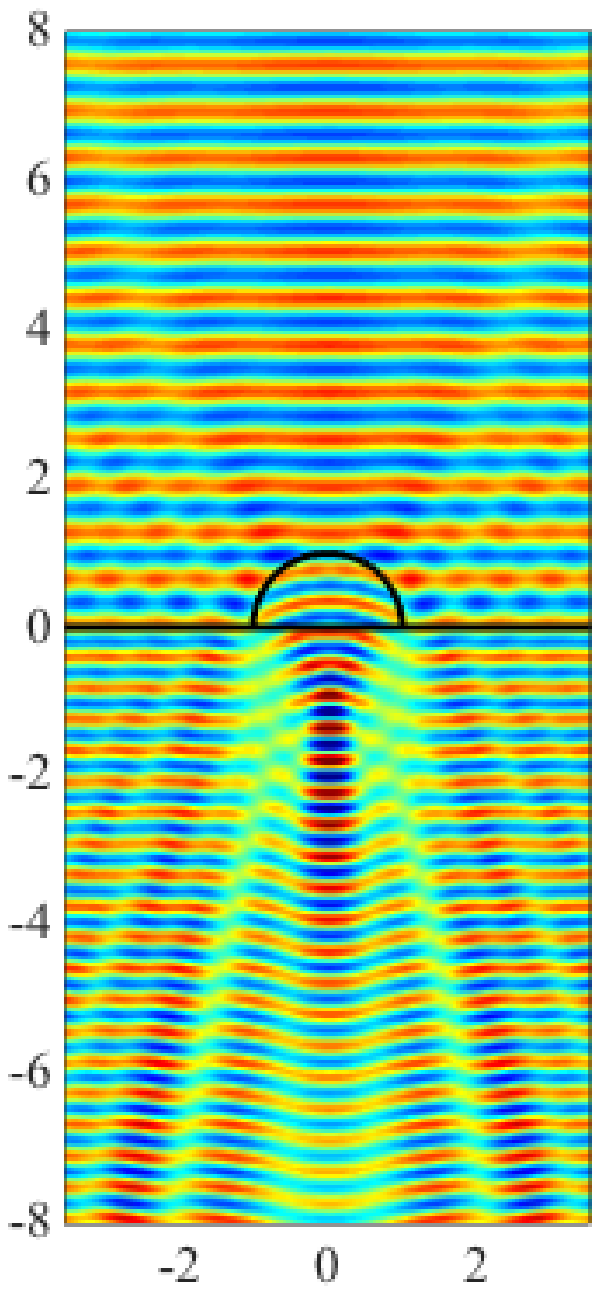}
        \caption{WGF method.}
        \label{fig:WGFM_pi_6}
    \end{subfigure}
    \hfill
    \begin{subfigure}[b]{0.3\textwidth}
        \centering
        \includegraphics[scale=0.45]{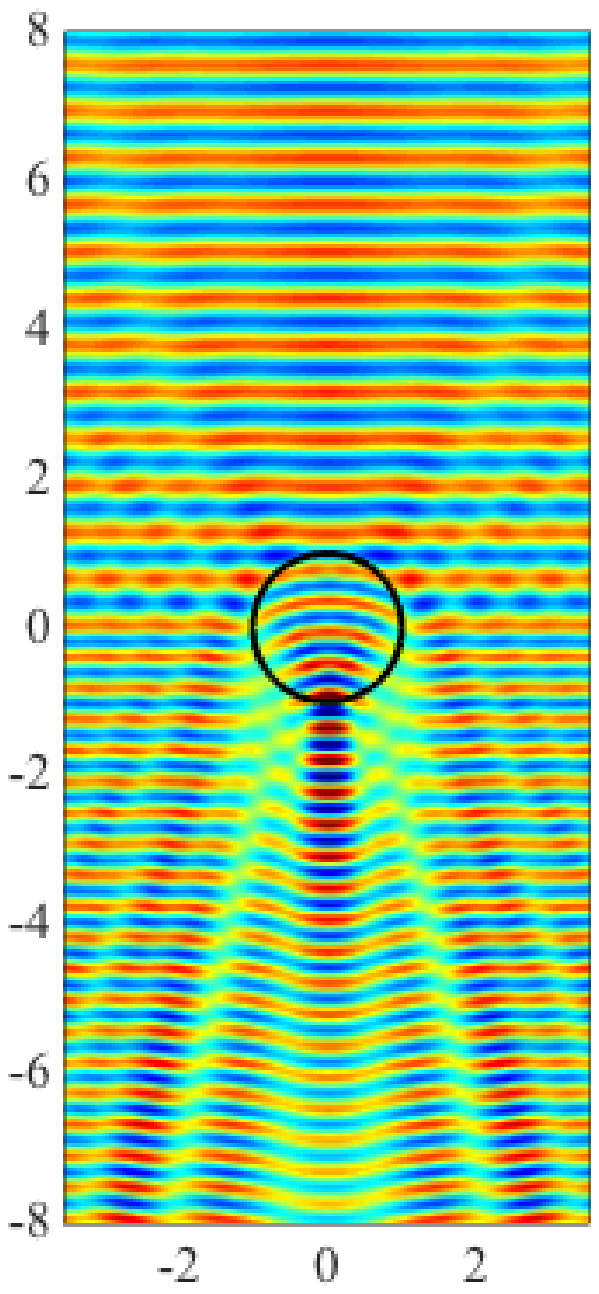}
        \caption{LGF method.}
        \label{fig:LGFM_pi_6}
    \end{subfigure}
    \hfill
    \begin{subfigure}[b]{0.3\textwidth}
        \centering
        \includegraphics[scale=0.45]{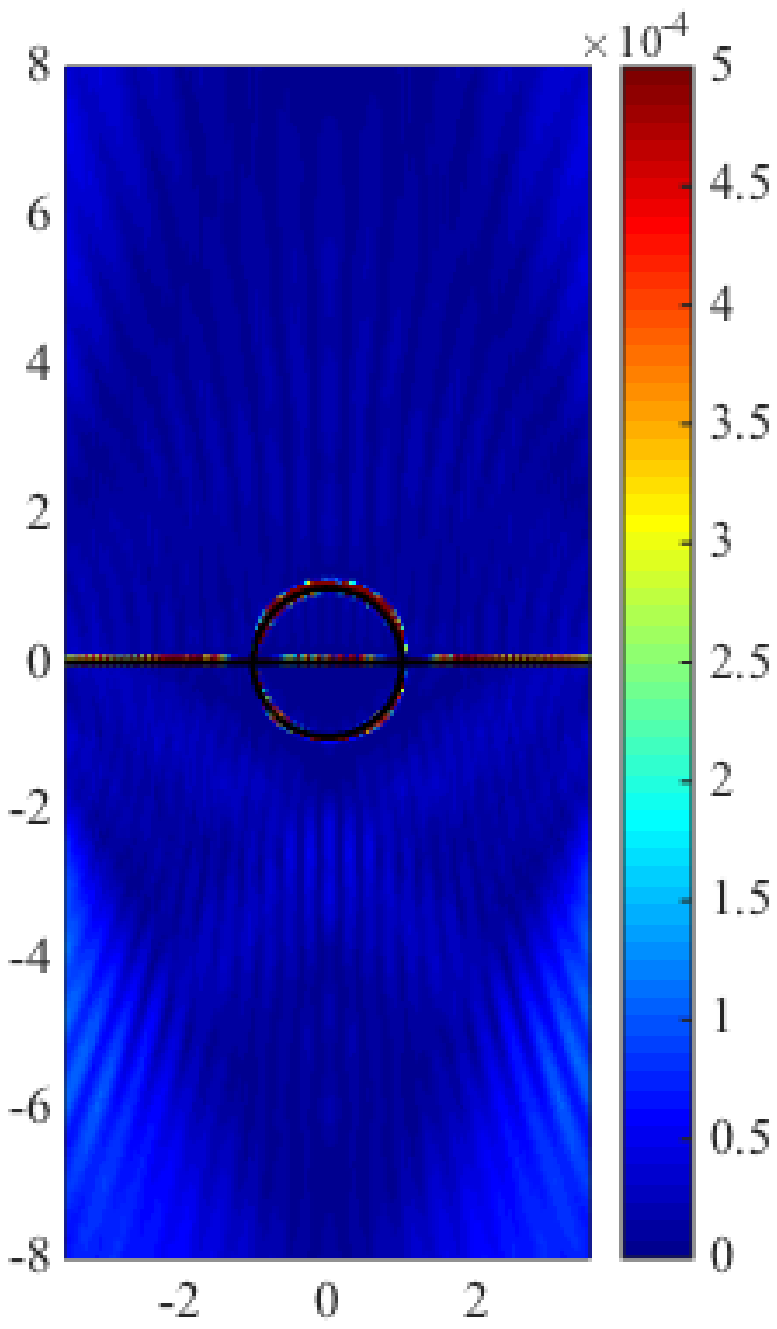}
        \caption{Difference.}
        \label{fig:pi_6_error}
    \end{subfigure}
    \caption{Real part of the total fields produced by the WGF method
   (first column) and the layer-Green-function method~\cite{PerezArancibia:2014fg}
   (second column), and  absolute value of the difference of the fields computed using the WGF method and layer-Green-function method  (third column) for the problem of scattering of plane-wave by a
   semi-circular bump for  $\alpha = -\pi/2$ (first row) and $\alpha=-\pi/6$ (second row) incidences. The width of the support of the selected window function is $2A=16\lambda\approx  10.053$ in all these calculations.
 The black lines represent the domains of the
   respective integral equation formulations.}
    \label{fig:total_field}
\end{figure}

\begin{figure}[ht!]
  \centering
    \includegraphics[scale=1.0]{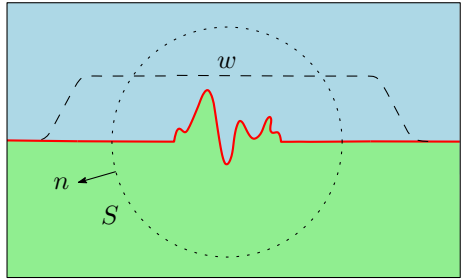}
    \caption{Surface $S$ utilized in \eqref{eq:scat_rep_formula}.}
  \label{fig:far_field_depic}
\end{figure}

Figure~\ref{fig:total_field} compares the total field obtained by
means of the WGF method and the layer-Green-function
method~\cite{PerezArancibia:2014fg} for the solution of the problem of
scattering of a plane-wave by a semi-circular bump of radius $a=1$ in
TE-polarization for wavenumbers $k_1=10$ and $k_2=15$ for
$\alpha=-\pi/2$ and $\alpha=-\pi/6$ incidences. The WGF solution, in
particular, was obtained from the solution of the integral
equation~\eqref{eq:IE_2nd_version} followed by evaluation of field
values on the basis
of~\eqref{eq:evaluation_formulae}. Figures~\ref{fig:pi_2_error}
and~\ref{fig:pi_6_error}, which display the absolute value of the
difference of the total fields computed using the WGF method and the
layer-Green-function method on a bounded portion of the strip
$\{w=1\}$ demonstrate the accuracy of the computed solutions in the
near field.
 


As may be expected, however, formulae~\eqref{eq:evaluation_formulae}
do not generally provide an accurate approximation of either far
fields or near fields outside a neighborhood of $\Gamma_A$. In order
to tackle this difficulty we consider the boundary $S$ of the disc $D$
mentioned above and depicted in~Figure~\ref{fig:far_field_depic}: $S$
encloses the portion of~$\Gamma$ that differs from the flat
interface~$\Pi$ and, as indicated above, it lies within a fixed region
within which superalgebraic convergence of the fields $u_1$ and $u_2$
takes place.  Application of the Green identities, integrating over
the region exterior to $S$ and utilizing the layer Green function
leads to the following integral representation of scattered field $u^s
= u-u^f$:
\begin{equation}
u^s(\nex) = \int_{S}\left\{\frac{\p G^{1}_{2}}{\p n_{\ney}} (\nex,\ney)u^s(\ney)-G^{1}_{2}(\nex,\ney)\frac{\p u^s}{\p n}(\ney)\right\}\de s_{\ney}\label{eq:scat_rep_formula}
\end{equation}
outside the region enclosed by $S$, where $G_{2}^{1}$ denotes the
layer Green function for the Helmholtz equation with wavenumbers $k_1$
in $\{x_2\geq 0\}$ and $k_2$ in $\{x_2<0\}$ that satisfies homogeneous
transmission conditions on the flat interface $\Pi$ (see Appendix~\ref{app:greens_function}).  Note that the
scattered field $u^s$ and its normal derivative on $S$ can be computed
directly utilizing~\eqref{eq:evaluation_formulae} since by
construction $S$ lies inside the region
where~\eqref{eq:evaluation_formulae} provides an accurate
approximation of the total field~$u$.

\begin{figure}[ht!]
  \centering
    \includegraphics[scale=0.250]{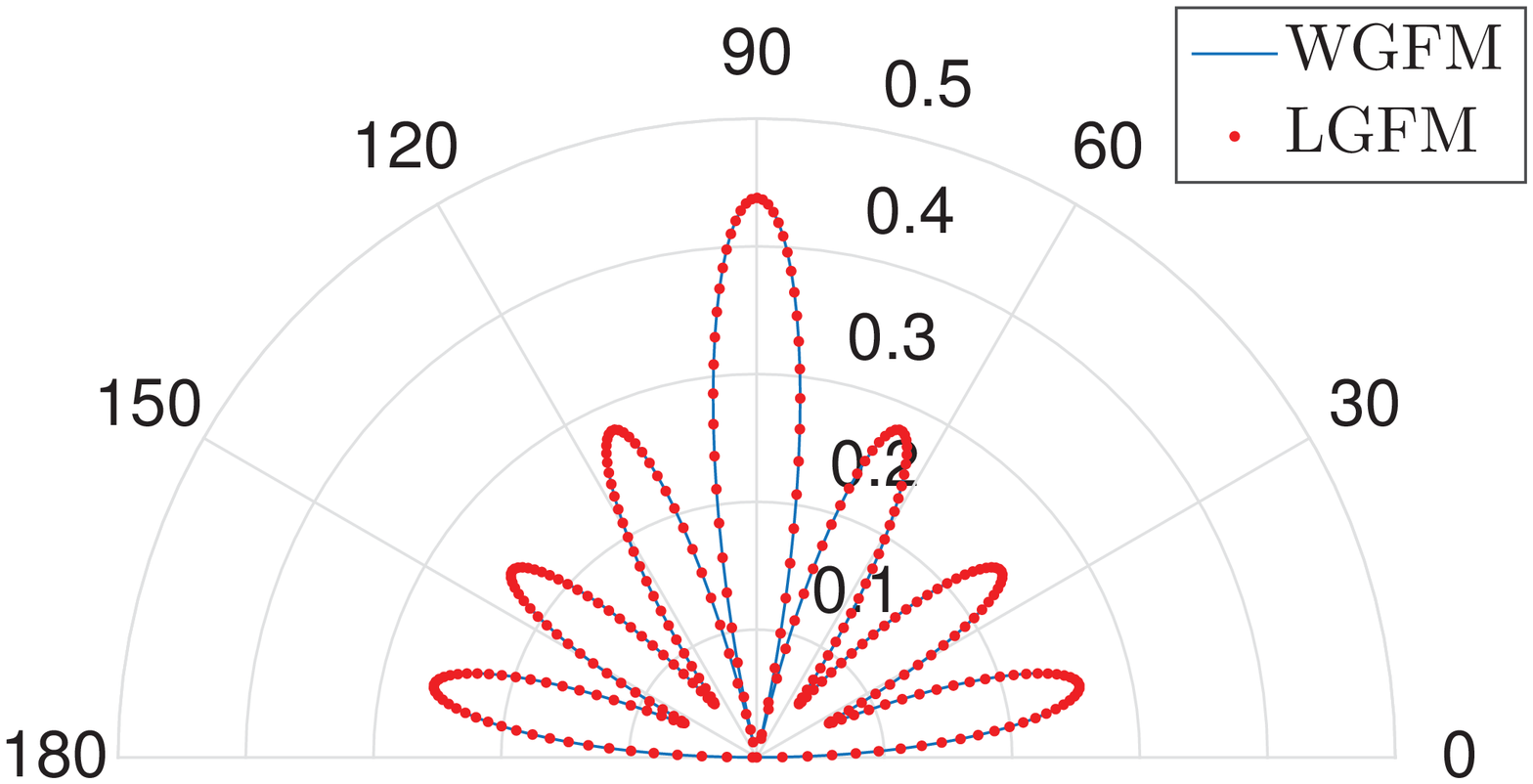}\hspace{0.5cm}
 \includegraphics[scale=0.250]{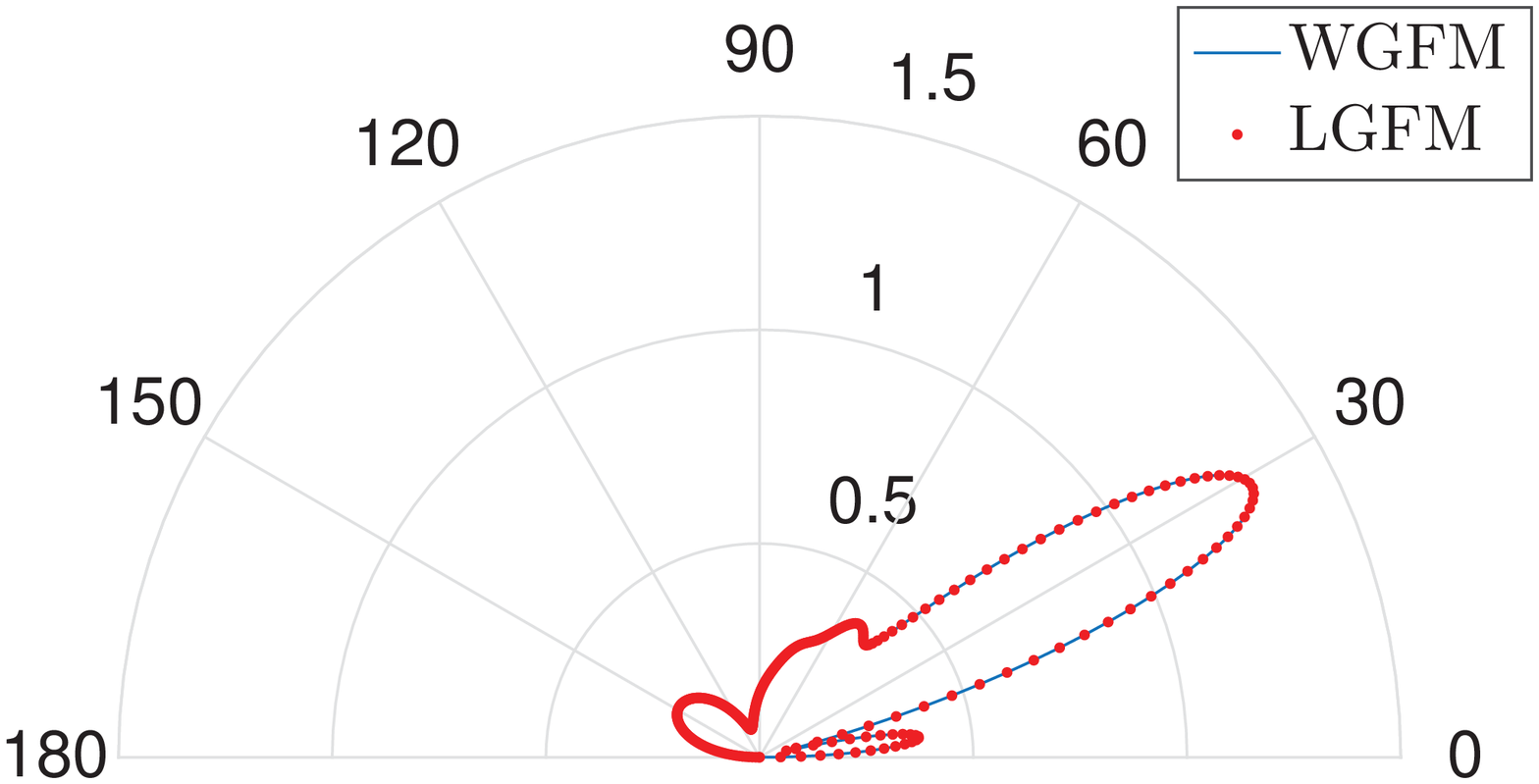}
  \caption{Far-field patterns obtained using the layer-Green-function method~\cite{PerezArancibia:2014fg} (red dotted curve) and the WGF method (continuous blue line) for the solution of the problem of scattering considered in this section at incidences $\alpha=-\pi/2$ (left) and $\alpha=-\pi/6$ (right) .}
  \label{fig:far_field}
\end{figure}

The far-field pattern $u_{\infty}(\hat \nex)$, which is related to the
scattered field by the asymptotic formula
$$
u^s(\nex) = \frac{\e^{ik_1 r}}{\sqrt{r}} u_{\infty}(\hat\nex) + \mathcal O(r^{-3/2}),\quad r=|\nex|\rightarrow\infty, \quad \hat\nex = \frac{\nex}{|\nex|},
$$
can be obtained from~\eqref{eq:scat_rep_formula} in a straightforward
manner by replacing $G_{2}^{1}$ by its asymptotic expansion as
$|\nex|\rightarrow\infty$.  The first order term of the asymptotic
expansion of the Sommerfeld integrals $\Phi_1$ and~$\Phi_2$~(equation~\ref{eq:somm_up}) in a given direction $\hat\nex =
(\cos\alpha,\sin\alpha)$, $0<\alpha<\pi$ can be obtained by the method of steepest descent by taking into account the contribution of the saddle point~\cite{Cui:1998fw} (branch point singularities and poles do not contribute to the first term of the asymptotic expansion of the two-layer Green function). Substitution of the result in
equation~\eqref{eq:scat_rep_formula} gives rise to the expression
 \begin{equation}
 u_{\infty}(\hat\nex) = \int_S\left\{\frac{\p H}{\p n_{\ney}}(\hat\nex,\ney) u^{s}(\ney)-H(\hat\nex,\ney)\frac{\p u^s}{\p n}(\ney)\right\}\de s_{\ney}\label{eq:far_field_integral}
\end{equation}
 for the far field $u_{\infty}(\hat\nex)$,  where
\begin{subequations}
\begin{align} 
H\left(\hat\nex,\ney\right) =&\frac{\nu(k_2^2-k_1^2)}{\sqrt{2\pi k_1 } (1+\nu)} \frac{\e^{-ik_1\hat\nex\cdot\ney}\e^{-2y_2\eta_1+i\pi/4}}{\left(\eta_2+\eta_1\right)\left(\eta_1+\nu\eta_2\right)}
+\frac{\e^{-i k_1 \hat\nex\cdot\ney+i\pi/4}}{\sqrt{8\pi k_1}}+\left(\frac{1-\nu}{1+\nu}\right)\frac{\e^{-i k_1 \hat{\bar\nex}\cdot \ney +i \pi/4}}{\sqrt{8\pi k_1}}
\end{align}
for $\ney\in \{y_2\geq 0\}$ and
\begin{equation}
 H\left(\hat\nex,\ney\right) =\frac{\nu k_1}{\sqrt{2\pi k_1 }} \frac{\sin(\alpha-\beta)\e^{-ik_1\hat\nex\cdot\ney}\e^{y_2(\eta_2-\eta_1)-i\pi/4}}{\eta_1+\nu\eta_2}
\end{equation}\label{eq:GF_far_field}\end{subequations}
for $\ney\in\{y_2<0\}$, where $\hat{\bar\nex} =\bar\nex/|\nex|=
(\cos\alpha,-\sin\alpha)$, $\ney = |\ney|(\cos\beta,\sin\beta)$,
$\eta_1 = \gamma_1(k_1\cos(\alpha-\beta))$ and
$\eta_2=\gamma_2(k_1\cos(\alpha-\beta))$ (see Appendix~\ref{app:greens_function} for the definition of $\gamma_1$ and $\gamma_2$).  Thus, unlike the
layer Green function $G_2^1$ itself, the far field associated with
$G_2^1$ can be computed inexpensively by means of the explicit
expressions~\eqref{eq:GF_far_field}. Figure~\ref{fig:far_field} provides a
comparison of the far-field patterns computed using the
layer-Green-function method and the WGF method proposed in this paper
for the example problem considered above in the present
section~\ref{field_eval}.

In view of this discussion, equations~\eqref{eq:evaluation_formulae}
and~\eqref{eq:scat_rep_formula} can be used to accurately and
efficiently evaluate near-fields and far-fields, respectively. These
are typically the quantities of interest in scattering simulations
involving layered media. The evaluation of the fields in an
intermediate region, such as a domain outside the neighborhood of
$\Gamma_A$ where~\eqref{eq:evaluation_formulae} yields an accurate
approximation, can also be approximated efficiently on the basis of
equation~\eqref{eq:scat_rep_formula}. Indeed, in such cases, for which
source points $\ney$ lie on $S$ and observation points $\nex$ are at a
certain distance away from $S$, the Sommerfeld
integrals~\eqref{eq:somm_up} and~\eqref{eq:somm_lw} (which contain
highly oscillatory and/or exponentially decaying integrands) can be
obtained by means of asymptotic numerical
methods~\cite{Asheim:2010kc,Bruno:2004tf} based on localization around
critical points~\cite{Cui:1998fw,PerezArancibia:2014fg}. 
\section{Numerical Experiments}\label{sec:num_exp}

This section illustrates the proposed methodology with a variety of
numerical results concerning dielectric and conducting media,
including relevant efficiency and accuracy studies. 

\begin{figure}[ht!]
  \centering
 \includegraphics[scale=0.5]{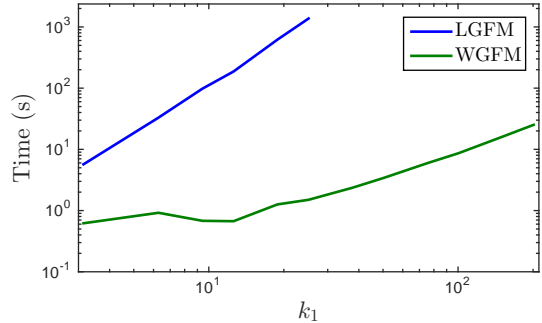}
  \caption{Computing times required by the WGF method (green line) and the layer-Green-function method~\cite{PerezArancibia:2014fg} (blue line) to create the linear systems of equations resulting from the Nystr\"om discretization of the relevant integral equations. }
  \label{fig:bump_eff}
\end{figure}

In our first example we consider once again the configuration
associated with Figure~\ref{fig:window_method_only} (i.e. the problem
of scattering by a semi-circular bump defect on a dielectric plane in
TE-polarization). Here we compare the computing times required to
create the systems of equations (which is the operation that dominates
the computing time in all the examples considered) that stem from the
discretization of the relevant integral equations by means of the WGF
method~\eqref{eq:IE_2nd_version} and the layer-Green-function
method~\cite[Eq. 7]{PerezArancibia:2014fg}. Figure~\ref{fig:bump_eff}
displays the computing times for various wavenumbers $k_1$ and
$k_2=2k_1$ for each method. The discretization density was held
proportional to $k_1$ to properly resolve the oscillatory character of
the integrands and the same discretization was used for both methods
on the bump, allowing for a point by point comparison of the
solutions. In all these examples the WGF method was optimized to
produce a maximum error of approximately $5\times 10^{-5}$ in the
computation of the density $\phi^w$ on the surface of the bump.
Similarly, the key parameters in the implementation of
layer-Green-function method (including the parameters associated to
the numerical evaluation of the Sommerfeld integrals) were adjusted to
yield the fastest possible solution within an error
of~$5\times10^{-5}$. Note that the last data points around
$k_1=8\pi\approx 25.1$ in Figure~\ref{fig:bump_eff} (which is the last
data point presented for the layer-Green-function method) shows that,
for such frequencies the WGF is approximately three orders of
magnitude faster than the layer-Green-function
method~\cite{PerezArancibia:2014fg}.

\begin{figure}[ht!]
  \centering
 \includegraphics[scale=0.65]{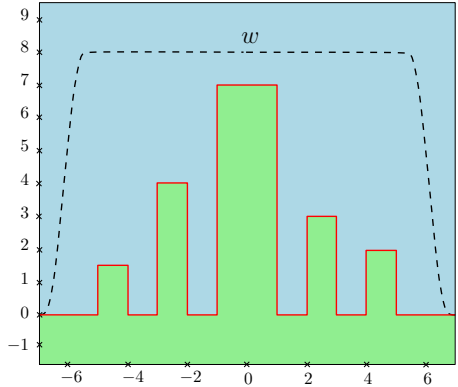}
 \caption{City-like geometry and windowing function used.}
  \label{fig:city}
\end{figure}

The problem of scattering by the city-like structure depicted in
Figure~\ref{fig:city} is considered next.  Figure~\ref{fig:city} also
displays the window function utilized in this example, which has been
amplified by a factor~8 for visualization purposes. In contrast with
the results presented previously in this paper, the case of
TM-polarization is considered for this test. In order to properly
account for the singular behavior of the fields near corners, the
necessary graded meshes were generated utilizing the value $p=4$ in
the method described in~\cite{Kress:1990vm}.  Table~\ref{tb:city}
reports the computing times required to form the relevant system
matrices for both the WGF method and the layer-Green-function method.
Both solvers were optimized to produce a maximum error of
$5\times10^{-3}$ in the solutions of the integral equation, and the
same computational grids were utilized to discretize the buildings for
both methods.

Table~\ref{tb:city} compares the computing times required by the WGF
method and the layer-Green-function method for two values of $k_2$. In
particular we note that, not only is the new method much faster than
the previous approach, but also that the speed-up factor grows: a
speed up factor in the hundreds for the value $k_2=2\pi$ is doubled as
$k_2$ is itself doubled to the value $k_2 =4\pi$. Additionally,
application of the layer-Green-function method in this context
requires use of fictitious curves underneath each
building~\cite{PerezArancibia:2014fg} each one of which (curves) must
itself be discretized, while the WGF method requires discretization of
the ground between the buildings and in the region where the windowing
takes place.  In the present case the layer-Green-function method
produced a system of 2384 unknowns while the WGF method produced a
nearly identical sized system of 2406 unknowns. At higher frequencies,
the WGF method requires fewer unknowns than the layer-Green-function
method, since, as demonstrated in Table~\ref{tb:size_freq}, at higher
frequencies the width of the windowing function can be decreased while
maintaining accuracy.

\begin{table}[h!]
\begin{center}
\begin{tabular}{|c|c|c|c|c|}
\hline
$k_1$ & $k_2$ & LGFM time & WGFM time & ratio \\
\hline $\pi$ & $2\pi$ & 588 s.& 3.07 s.& 192\\
\hline $\pi$ & $4\pi$ & 3579 s.& 9.10 s.& 393\\
\hline
\end{tabular}
\caption{\label{tb:city} Computing times required by the
  layer-Green-function method and the WGF method to produce integral
  equation solutions with an accuracy better than $5\times10^{-3}$ for
  the city-like geometry displayed in Figure~\ref{fig:city}. }
\end{center}
\end{table}

\begin{table}[h!]
\begin{center}
\begin{tabular}{|c|c|c|}
\hline
$k_1$ & $k_2$ & $A$\\
\hline $\pi$ & $2\pi$ & 6.5\\
\hline $2\pi$ & $4\pi$ & 3.5\\
\hline $4\pi$ & $8\pi$ & 1.75\\
\hline $8\pi$ & $16\pi$ & 1.1875\\
\hline
\end{tabular}
\caption{\label{tb:size_freq} Extent of the windowed region required
  by the WGF method~\eqref{eq:IE_2nd_version} to maintain an accuracy
  of $5\times 10^{-5}$ in the approximation of the surface fields for
  the problem of scattering from a semi-circular bump of unit radius
  with various wavenumbers. The angle of incidence was taken to equal
  $\alpha=-\pi/8$ .}
\end{center}
\end{table}

\begin{figure}[ht!]
  \centering
  \includegraphics[scale=0.35]{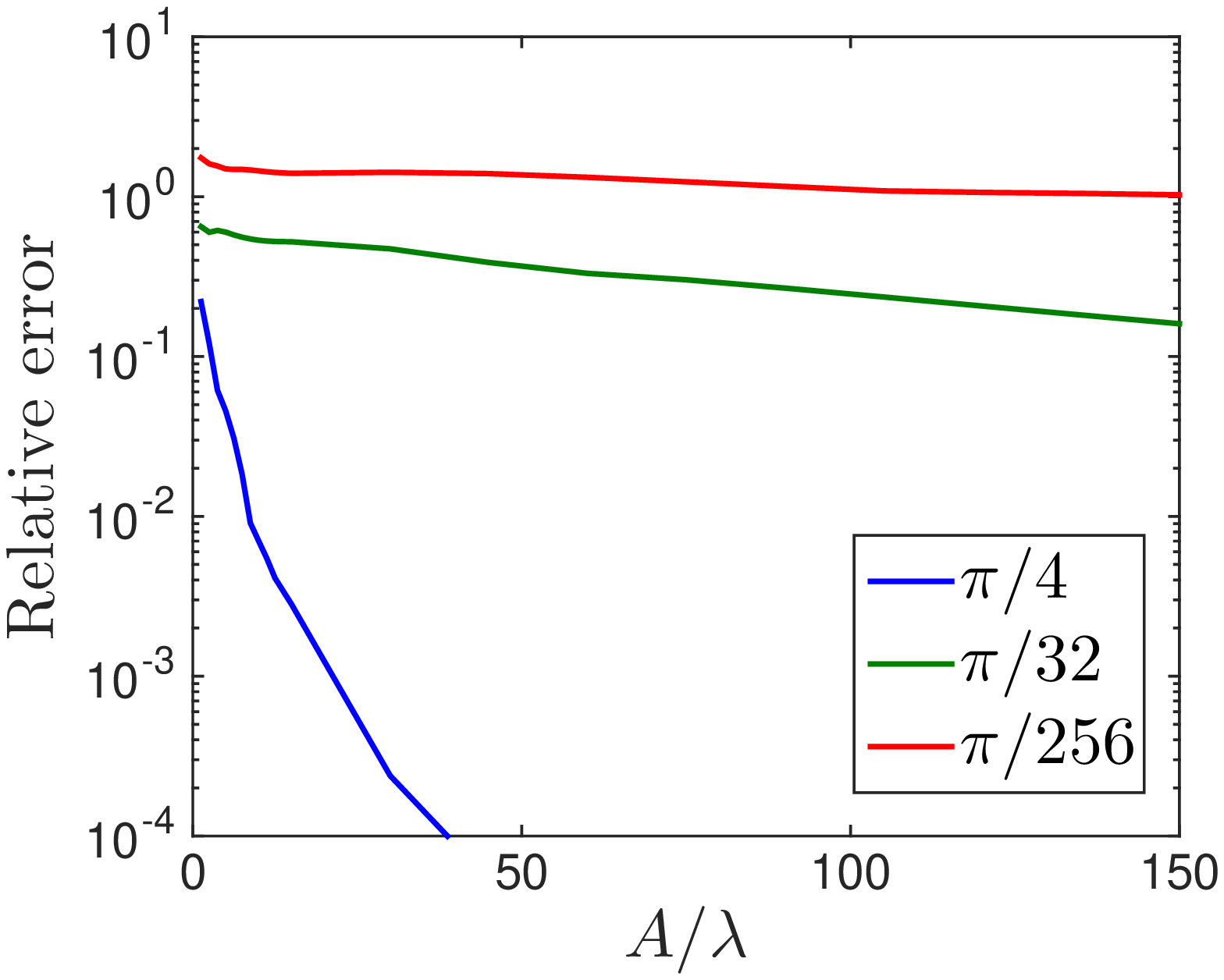}\hspace{0.5cm}
 \includegraphics[scale=0.35]{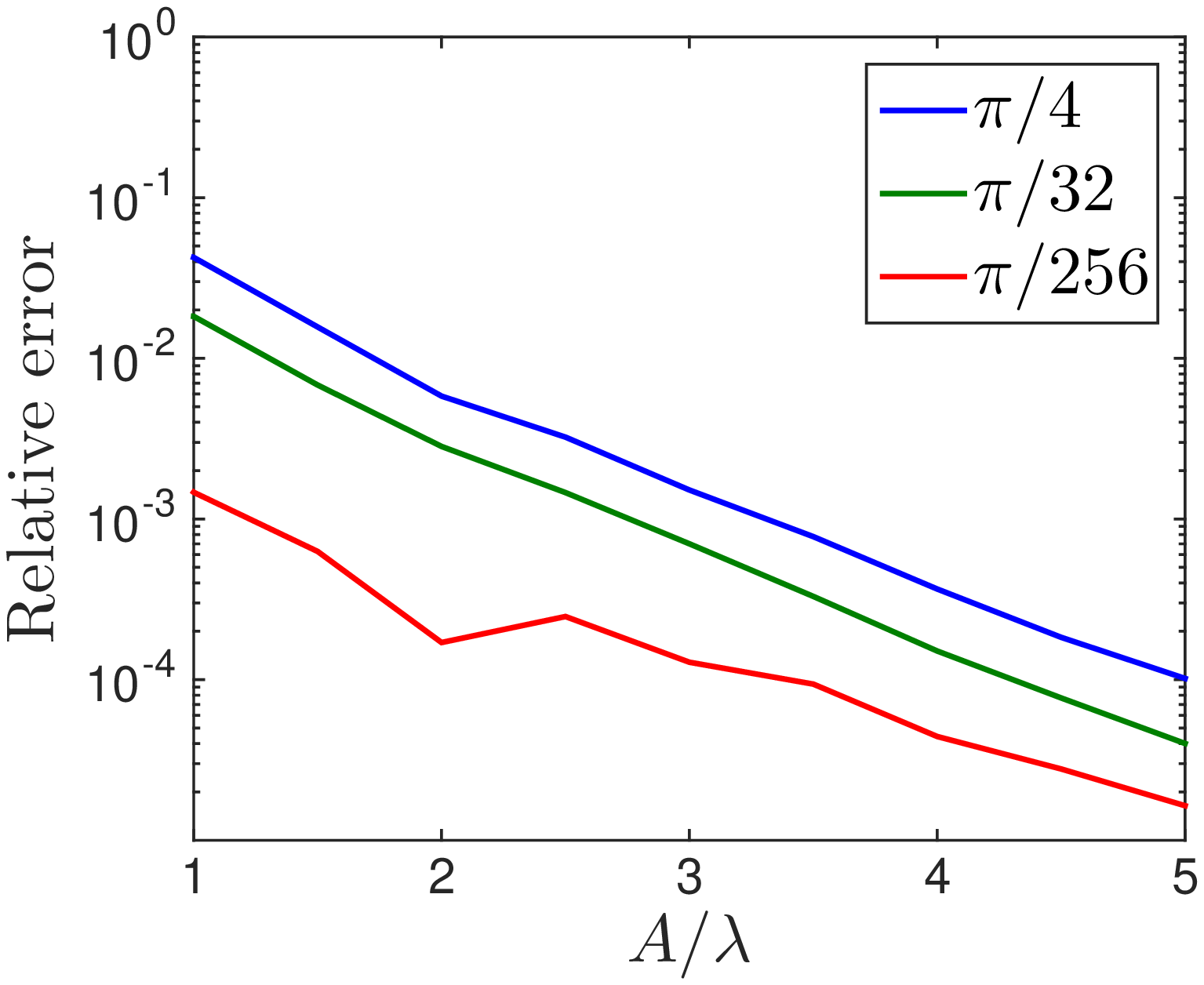}
 \caption{Errors in the integral densities resulting from numerical
   solution by means of the layer-Green-function
   method~\eqref{eq:angledep} (left) and the WGF
   method~\eqref{eq:IE_2nd_version} (right) for the city-like
   structure depicted in Figure~\ref{fig:city}, for various window sizes
   and angles of incidence---including extremely shallow
   incidences. Clearly, the WGF method computes integral densities
   with super-algebraically high accuracy uniformly for all angles of
   incidence.}
  \label{fig:absorb}
\end{figure}

As an additional example we consider once again the city-like
structure depicted in Figure~\ref{fig:city} but assuming an absorbing
media in the ground and buildings: here we thus take $k_1=2\pi$ and
$k_2=4\pi(1+i/100)$. Figure~\ref{fig:absorb} demonstrates the
convergence of both the naive windowing algorithm~\eqref{eq:angledep}
and the full WGF method~\eqref{eq:IE_2nd_version}. The advantages
provided by the full WGF approach can be appreciated clearly in this
figure: in the naive method convergence near grazing is extremely slow
while for the full WGF method the convergence is actually faster near
grazing than for non-grazing configurations. In particular, the WGF
method requires no more than 5 wavelengths of ground for a full four
digits of accuracy, independently of the incidence angle.
	
\begin{figure}[ht!]
  \centering
 \includegraphics[scale=0.3]{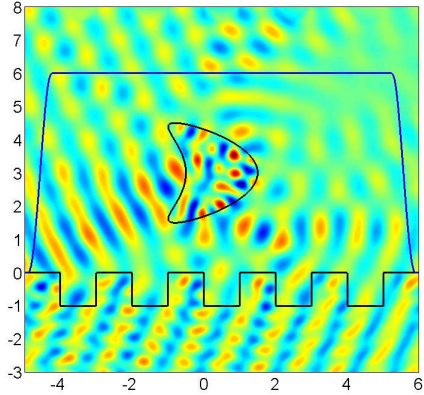}\
 \caption{Scattering geometry containing a kite structure above a
   finite rectangular grating in an otherwise undisturbed planar
   ground. A windowing function large enough to produce an error
   smaller than 1\% in the integral equation solution is shown along
   with the corresponding near fields; $k_1=2\pi$ and $k_2=4\pi$.}
  \label{fig:CKnear}
\end{figure}

\begin{figure}[ht!]
  \centering
 \includegraphics[scale=0.35]{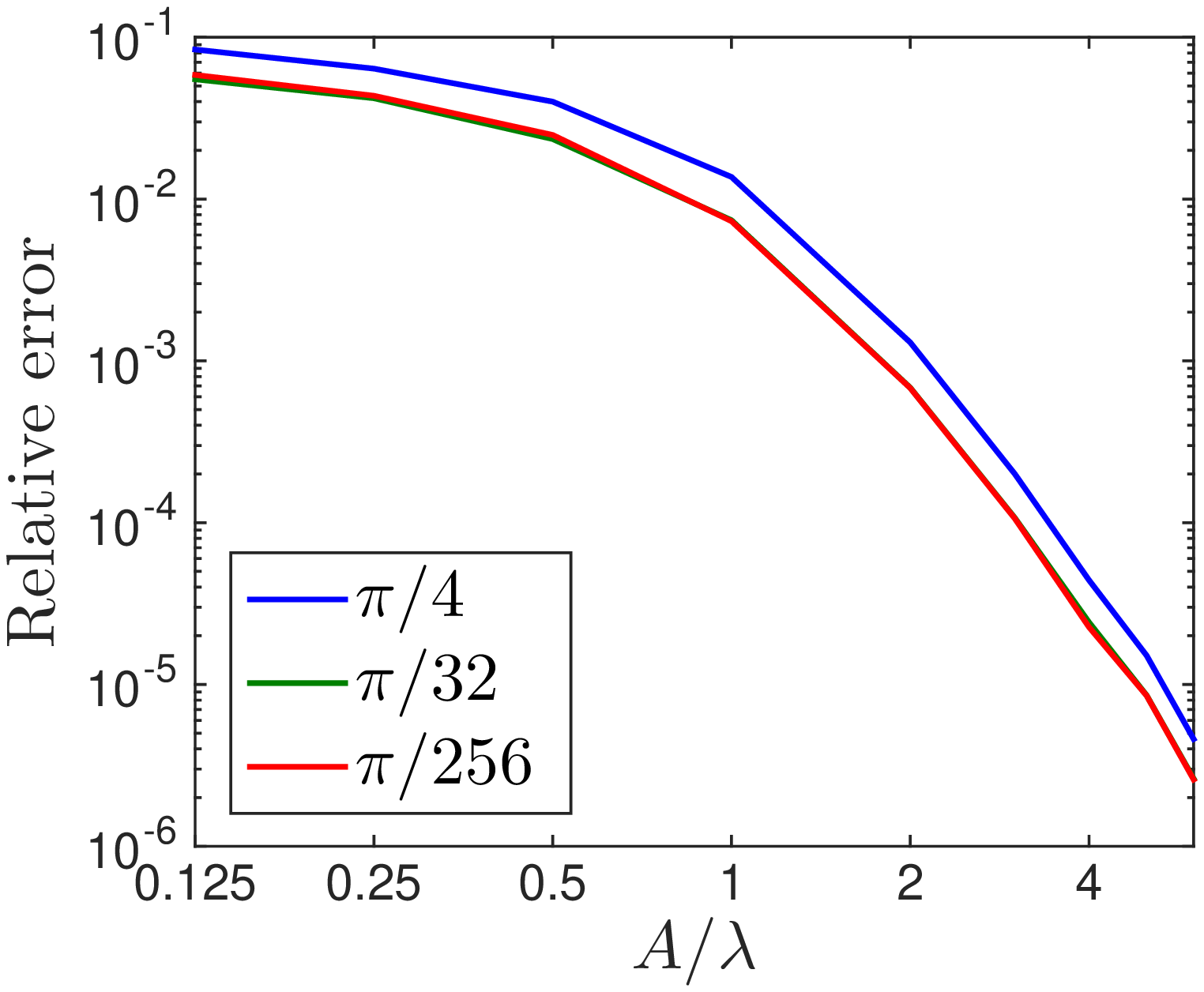}\hspace{0.5cm}
 \includegraphics[scale=0.35]{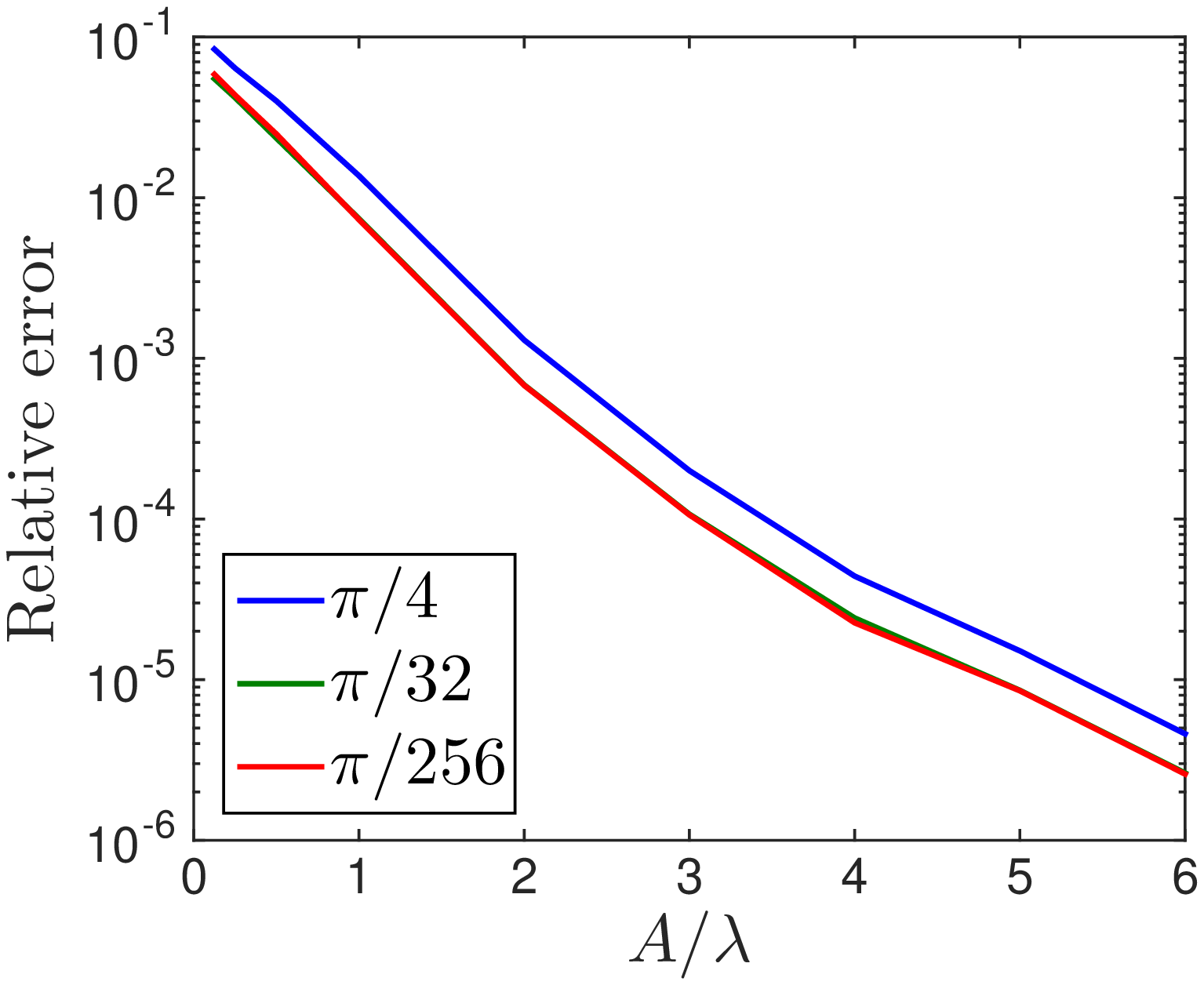}
 \caption{Errors in the integral densities resulting from numerical
   solution of~\eqref{eq:IE_2nd_version} for the structure depicted in
   Figure~\ref{fig:CKnear} by means of the full WGF method, for various
   window sizes and angles of incidence---including extremely shallow
   incidences. Left: log-log scale. Right: semi-log scale.  Once again
   we see that, the WGF method computes integral densities with
   super-algebraically high accuracy uniformly for all angles of
   incidence.}
  \label{fig:CKconv}
\end{figure}

For our last numerical example we consider an obstacle above the
ground, but not connected to it, with a finite number of indentations
under the ground level. Figure~\ref{fig:CKnear} displays the geometry
under consideration, together with a selection of window function
which yields an error of approximately 1\% in the integral equation
solution and corresponding near fields for a plane-wave illumination
with incidence angle equal to $\alpha=-\pi/8$ from the horizontal
under TE polarization. Once again, as demonstrated in
Figure~\ref{fig:CKconv} exponential convergence is observed as
$A/\lambda$ grows.

\section*{Acknowledgments}
The authors gratefully acknowledge
support from the Air Force Office of Scientific Research and the
National Science Foundation.

\appendix
\section{Green function for a two-layer medium\label{app:greens_function}}
Consider the Helmholtz equation in the regions
$\Omega_1=\{(x_1,x_2)\in\R^2,x_2>0\}$ and
$\Omega_2=\{(x_1,x_2)\in\R^2,x_2<0\}$ with respective wavenumbers 
$k_1$ and $k_2$. The Green function of the problem
satisfies:
\begin{equation}
\begin{array}{cccllll}
\Delta_{\nex} G+k_j^2 G&=&-\delta_{\ney}&\mbox{in}&\Omega_j,\smallskip\\
G|_{x_2=0^+}&=&G|_{x_2=0^-} &\mbox{on}& \{x_2=0\}, \smallskip\\
\displaystyle\frac{\p G}{\p x_2}\big|_{x_2=0^+} &=&\displaystyle\nu\frac{\p G}{\p x_2}\big|_{x_2=0^-} &\mbox{on}&\{x_2=0\},
\end{array}\label{eq:def_greens_function}
\end{equation}
and the Sommerfeld radiation condition at infinity, where~$\delta_{\ney}$ denotes the Dirac delta distribution supported at the
point~$\ney$. As is known $G$ can be computed explicitly in
terms of Sommerfeld integrals. To obtain such explicit expressions,
given a fixed point $\ney$ we define the functions
$\varphi_j(\nex)=G(\nex,\ney)$, $\nex\in\Omega_j$. Expressing
$\varphi_j$ as inverse Fourier transforms
\begin{equation}
\varphi_j(x_1,x_2) = \frac{1}{2\pi}\int_{-\infty}^{\infty} \widehat\varphi_j(\xi,x_2)\e^{i\xi (x_1-y_1)}\de \xi\label{eq:fourier_transform}
\end{equation} and replacing \eqref{eq:fourier_transform} in \eqref{eq:def_greens_function} a system of ordinary differential equations for the unknown functions $\widehat\varphi_j$ is obtained which can be solved analytically. 
Two cases arise. For $\ney\in \Omega_1$, the solution of the ODE system is given by 
\begin{equation}
\begin{split}
\widehat\varphi_1(\xi,x_2) =&\frac{\e^{-\gamma_1|x_2-y_2|}}{2\gamma_1} +\left(\frac{1-\nu}{1+\nu}\right)\frac{\e^{-\gamma_1|x_2+y_2|}}{2\gamma_1}\\
&+\frac{\nu(k^2_{2}-k^2_1)}{(\gamma_1+\nu\gamma_2)(1+\nu)}\frac{\e^{-\gamma_1 (x_2+y_2)}}{\gamma_1(\gamma_1+\gamma_2)},\\
\widehat\varphi_2(\xi,x_2) =&\frac{\e^{-\gamma_1(y_2-x_2)}}{(1+\nu)\gamma_1}+\ \left(\frac{\e^{\gamma_2x_2-\gamma_1y_2}}{\gamma_1+\nu\gamma_2}-\frac{\e^{-\gamma_1(y_2-x_2)}}{(1+\nu)\gamma_1}\right),
\end{split}
\end{equation}
where $\gamma_j = \sqrt{\xi^2-k_j^2}$.  The determination of
physically admissible branches of the functions $\gamma_j(\xi) =
\sqrt{\xi-k_j}\sqrt{\xi+k_j}$ require selection of branch cuts for
each one of the two associated square root functions. The relevant
branches are $-3\pi/2 \leq \arg(\xi-k_j)<\pi/2$ for
$\sqrt{\xi-k_j}$ and $-\pi/2\leq\arg(\xi+k_j)<3\pi/2$ for
$\sqrt{\xi+k_j}$.  Taking the inverse Fourier
transform~\eqref{eq:fourier_transform} of $\widehat\varphi_j$ and using
the identity
$$
\int_{-\infty}^{\infty}\frac{\e^{-\gamma_j|x_2-y_2|}}{4\pi \gamma_j}\e^{i\xi(x_1-y_1)}\de \xi = \frac{i}{4}H^{(1)}_0(k_j|\ney-\nex|),
$$
we obtain
\begin{equation}
\begin{split}
\varphi_1(\nex) =&\frac{i}{4} H_0^{(1)}(k_1|\nex-\ney|)+\frac{i}{4}\left(\frac{1-\nu}{1+\nu}\right)H_0^{(1)}(k_1|\overline{\nex}-\ney|)\\&+\Phi_1(\nex,\ney),\\
\varphi_2(\nex) =&\frac{i}{2}\frac{1}{1+\nu} H_0^{(1)}(k_1|\nex-\ney|)+\Phi_2(\nex,\ney),
\end{split}
\end{equation}
where the functions $\Phi_j$ are given by
\begin{equation}
\begin{split}
\Phi_1(\nex,\ney) =& \displaystyle\frac{\nu(k_2^2-k_1^2)}{\pi (1+\nu)} \displaystyle\int_{0}^{\infty}\frac{\e^{-\gamma_1(x_2+y_2)}\cos(\xi(x_1-y_1))}{\gamma_1(\gamma_2+\gamma_1)(\gamma_1+\nu\gamma_2)}\de \xi,\bigskip\\
\Phi_2(\nex,\ney) =& \displaystyle\frac{1}{\pi}\int_{0}^{\infty}\left(\frac{\e^{\gamma_2 x_2-\gamma_1 y_2}}{\gamma_1+\nu\gamma_2}-\frac{\e^{\gamma_1(x_2-y_2)}}{(1+\nu)\gamma_1}\right)\cos(\xi(x_1-y_1))\de \xi,
\end{split}\label{eq:somm_up}
\end{equation} 
Similarly, the solution of the ODE system for $\ney\in\Omega_2$ is given by 
\begin{equation*}
\begin{split}
\widehat\varphi_1(\xi,x_2) =&\frac{\nu\e^{-\gamma_2(x_2-y_2)}}{(1+\nu)\gamma_2}+  \left(\frac{\nu\e^{-\gamma_1x_2+\gamma_2y_2}}{\gamma_1+\nu\gamma_2}-\frac{\nu\e^{-\gamma_2(x_2-y_2)}}{(1+\nu)\gamma_2}\right),\\
\widehat\varphi_2(\xi,x_2) =&\frac{\e^{-\gamma_2|x_2-y_2|}}{2\gamma_2} +\left(\frac{\nu-1}{\nu+1}\right)\frac{\e^{-\gamma_2|x_2+y_2|}}{2\gamma_2}\\
&+\frac{\nu(k^2_1-k^2_2)\e^{\gamma_2(x_2+y_2)}}{(\gamma_1+\nu\gamma_2)(1+\nu)\gamma_2(\gamma_2+\gamma_1)}.
\end{split}
\end{equation*}

Taking inverse Fourier transform \eqref{eq:fourier_transform} we now obtain 
\begin{equation}
\begin{split}
\varphi_1(\nex) =&\frac{i}{2}\frac{\nu}{1+\nu} H_0^{(1)}(k_2|\nex-\ney|)+\Psi_1(\nex,\ney),\\
\varphi_2(\nex) =&\frac{i}{4} H_0^{(1)}(k_2|\nex-\ney|)+\frac{i}{4}\left(\frac{\nu-1}{\nu+1}\right)H_0^{(1)}(k_2|\overline{\nex}-\ney|)\\
&+\Psi_2(\nex,\ney),
\end{split}
\end{equation}
where the functions $\Psi_j$ are given by
\begin{equation}
\begin{split}
\Psi_1(\nex,\ney) =& \frac{\nu}{\pi}\int_{0}^{\infty}\!\!\left(\!\!\frac{\e^{\gamma_2 y_2-\gamma_1 x_2}}{\gamma_1+\nu\gamma_2}-\frac{\e^{-\gamma_2(x_2-y_2)}}{(1+\nu)\gamma_2}\right)\cos(\xi(x_1-y_1))\de \xi,\\
\Psi_2(\nex,\ney) =& \frac{\nu(k_1^2-k_2^2)}{\pi (1+\nu)}\int_{0}^{\infty}\frac{\e^{\gamma_2(x_2+y_2)}\cos(\xi(x_1-y_1))}{\gamma_2(\gamma_1+\gamma_2)(\gamma_1+\nu\gamma_2)}\de \xi.
\end{split}\label{eq:somm_lw}
\end{equation}
The gradient of the Green function is evaluated from the expressions
above by differentiation under the integral sign.

\bibliographystyle{abbrv}
\bibliography{paper_SIAM_final}

%

\end{document}